\documentclass[aps,pre,twocolumn,nofootinbib,tightenlines,superscriptaddress,amsmath,amssymb,final,letterpaper]{revtex4-1}
\usepackage[utf8]{inputenc}
\usepackage{graphicx}
\usepackage{amsmath,amssymb,amsthm}
\usepackage{natbib}
\usepackage{bbold}
\usepackage{calc}
\usepackage{bm}
\usepackage{color}
\usepackage{hyperref}
\usepackage{multirow}
\usepackage[dvipsnames]{xcolor}

\hypersetup{bookmarks=true, unicode=false, pdftoolbar=true, pdfmenubar=true, pdffitwindow=false, pdfstartview={FitH}, pdfnewwindow=true, colorlinks=true, linkcolor=blue, citecolor=blue, filecolor=magenta, urlcolor=blue}

\graphicspath{{./figs/}} 

\newcommand{\etal}{{\it{}et~al.}}

\newcommand{\Ord}{\mathrm{O}}

\begin{document}
\title{Bayesian inference of network structure from unreliable data}
\author{Jean-Gabriel Young}
\affiliation{Center for the Study of Complex Systems, University of Michigan, Ann Arbor, Michigan 48109, USA}
\author{George T. Cantwell}
\affiliation{Department of Physics, University of Michigan, Ann Arbor, Michigan 48109, USA}
\author{M. E. J. Newman}
\affiliation{Center for the Study of Complex Systems, University of Michigan, Ann Arbor, Michigan 48109, USA}
\affiliation{Department of Physics, University of Michigan, Ann Arbor, Michigan 48109, USA}

\begin{abstract}
  Most empirical studies of complex networks do not return direct, error-free measurements of network structure.  Instead, they typically rely on indirect measurements that are often error-prone and unreliable.  A fundamental problem in empirical network science is how to make the best possible estimates of network structure given such unreliable data.  In this paper we describe a fully Bayesian method for reconstructing networks from observational data in any format, even when the data contain substantial measurement error and when the nature and magnitude of that error is unknown.  The method is introduced through pedagogical case studies using real-world example networks, and specifically tailored to allow straightforward, computationally efficient implementation with a minimum of technical input.  Computer code implementing the method is publicly available.
\end{abstract}

\maketitle

\section{Introduction}

Networks are widely used as a convenient mathematical representation of the relationships between elements of a complex system~\cite{Newman18c}.  Network methods have been fruitfully applied to aid our understanding of systems in physics, biology, computer and information sciences, the social sciences, and many other areas.  A typical empirical network study will first determine the structure of a network of interest, using experiments, field observations, or archival data, then analyze that structure to reveal features of interest, using any of the many quantitative analysis techniques developed for this purpose \cite{kolaczyk2009statistical}.

In this paper we focus on the first part of this process, on how we determine the structure of a network from appropriate empirical observations.  At first sight, this appears to be a straightforward task.  Most studies aim to measure the presence or absence of edges in a network, either singly or in some collective fashion, and then assemble a picture of the complete network by aggregating many such measurements.  Upon further consideration, however, it is apparent that the situation is not so simple, because most network measurements do not tell us unambiguously about the presence or absence of edges.  At best, they give us a noisy indication, and in many cases they merely hint obliquely at the network structure~\cite{newman2018network}.

As an example, consider a protein-protein interaction network representing the pattern of physical interactions between proteins in the cell.  Such networks can be measured in a relatively direct manner.  Techniques such as affinity purification and two-hybrid screens can be used to determine whether a given protein interacts with others~\cite{Ito00_short,Krogan06_short}.  These techniques are notoriously unreliable, however, returning high rates of both false positives and false negatives, so that individual measurements do not themselves tell us the structure of the network~\cite{sprinzak2003reliable}.  Measurements may have to be repeated many times in order to separate the signal from the noise~\cite{Rolland14_short}.

A more complex example is the measurement of a social network such as a network of who is friends with whom.  Such networks can be measured by conducting surveys in which people are asked to identify their friends~\cite{WF94}.  Here, however, things can get complicated.  For instance, it happens often that person~A identifies person~B as a friend, but person~B does not identify person~A.  In some communities fully a half of all friendships are ``one-way'' friendships of this kind~\cite{VK08,BN13}.  Should two such people be considered friends?  Why do they disagree?  Is one of the individuals mistaken, or forgetful, or not telling the truth?  Perhaps the two are using different definitions of friendship?  Things become harder still when we consider other data that are commonly gathered in such surveys, such as age, gender, race, occupation, income, or educational level of participants.  Friendships are well known to depend strongly on such personal characteristics and in cases where we are uncertain about a hypothesized friendship between two people we may be able to use their traits to make a better informed decision about whether they are really friends~\cite{MSC01}.  The best estimate of whether two people are friends may thus be the result of a complex calculation that combines many inputs.

Most empirical studies of network structure, however, do not take such an approach.  Even though network data are known to be noisy and imperfect \cite{holland1973structural,sprinzak2003reliable,amini2004issues,whitehead2008analyzing,Sporns10,wang2012measurement,wiese2015you}, many experimenters nonetheless rely on simple direct measurements of the presence of edges and effectively make the assumption that the resulting data \emph{are} the network.  In a protein-protein interaction network they assume an interaction to be present if, for example, it is observed to exist enough times when measured.  In a friendship network two people are assumed to be friends if one identifies the other as a friend.  And yet it is clear from the discussion above that the true situation is more complicated than this.  In realistic circumstances, we have a heterogeneous body of data, potentially of many different types, potentially unreliable, and varying in its relationship to the actual network structure.  In such circumstances, traditional methods for inferring network structure from data may be inadequate and in some cases outright misleading.

In this paper we present a general framework for inferring network structure from measurements using methods of Bayesian inference, along with a complete software pipeline implementing that framework.  The methods we describe take empirical measurements of a system and return a posterior distribution over possible network structures, thereby telling us not only what the likely structure is but also giving us an estimate of our certainty about that structure.  Our approach requires a minimum of input from the experimenter, the only information they need supply (other than the data) being a description of how the measurements depend on the underlying network structure, which is specified at a high level in the form of probability distributions---we give a number of examples in this paper.  The rest of the calculation, from data to posterior distribution and final network structure, is performed automatically.  Application of Bayesian inference methods often requires experimentation to find the best approach for a particular application.  The level of automation in our framework makes this a straightforward task, allowing one to easily experiment with different strategies and quickly see the results~\cite{gelman2013philosophy}.

The problem we address falls in the general area of \emph{network reconstruction}, the problem of determining the structure of a network from the available data.  There has been a considerable amount of previous work on network reconstruction, much of it in the subject-specific literature and directed at the reconstruction of particular types of networks or using particular types of data.  Methods have been proposed, for example, for geographic co-location data~\cite{eagle2006reality,eagle2009inferring,crandall2010inferring,cranshaw2010bridging}, social networks~\cite{butts2003network}, brain scans~\cite{le2018estimating,tang2018connectome,wang2019common}, and biochemical networks~\cite{jansen2003bayesian,Krogan06_short,jiang2012latent,birlutiu2014bayesian}.  Perhaps the closest precursor to our work is that of Butts~\cite{butts2003network}, who developed Bayesian methods and Gibbs sampling techniques for estimating social network structure from unreliable social surveys.  Priebe and collaborators~\cite{priebe2015statistical} have developed similar ideas, incorporating structured priors over networks to improve inference, an approach that has been echoed in statistics~\cite{le2018estimating,tang2018connectome,wang2019common} and network science~\cite{peixoto2018reconstructing}.  A key difference between these approaches and our own, however, is that they were developed with particular measurement processes in mind, whereas our methods can accommodate almost arbitrary ones.  The approach we propose also differs from our own related previous work~\cite{newman2018network,newman2018estimating,young2019reconstruction} in putting forward an estimation algorithm that works essentially automatically with arbitrary models, using ideas borrowed from the literature on finite mixture models~\cite{titterington1985statistical,mclachlan2004finite}.

Looking more broadly, there is also a considerable volume of work that tackles other problems concerning the reliability of network data, many of which could also be addressed using variants or extensions of the methods proposed here.  The problem of \emph{link prediction}, for instance, involves predicting missing edges in partially observed networks and a range of algorithms have been proposed that achieve good performance~\cite{LK07,CMN08,guimera2009missing,huisman2009imputation,KL11}.  Also much studied is the problem of inferring network structure from non-network data~\cite{brugere2018network}, such as time-series~\cite{li2008mining}, gene expression data~\cite{bansal2007infer}, the spread of information or disease~\cite{gomez2012inferring,NS12}, and various local network features and node properties~\cite{squartini2017maximum,yuan2007model}.  A separate literature deals with problems of \emph{network sampling}, addressing how choices such as which nodes or edges we measure can affect our estimates of network properties and how best to make those choices~\cite{kolaczyk2009statistical,orbanz2017subsampling,stumpf2005subnets,lee2006statistical}.  \emph{Disambiguation} or \emph{entity resolution} is the process of accurately identifying the nodes in a network in the presence of potential sources of error such as duplicate nodes or nodes that have been inadvertently combined~\cite{butts2009revisiting,FGL12,wang2012measurement}.  Some methods have also been proposed that aim to perform more than one of these tasks at once, such as simultaneous link prediction and disambiguation~\cite{namata2016collective}.  Finally, there is also a growing literature on making best estimates of derived network metrics starting from probabilistic representations of network structure~\cite{pfeiffer2011methods,bonchi2014core,MBN16,poisot2016structure,khan2018uncertain}.  These methods use the kinds of structural estimates we develop in this paper as input to calculations of further quantities of interest, such as centrality measures, path lengths, correlations, community structure, or core decompositions~\cite{MBN16,bonchi2014core}.

This manuscript is organized as follows.  In Section~\ref{sec:overview} we first give an overview of the framework we propose for inference of network structure from unreliable data.  Then in Section~\ref{sec:case_studies} we describe in depth two example applications, illustrating the construction of appropriate measurement models and the practical implementation of the method.  In Section~\ref{sec:concs} we give our conclusions and suggest some directions for future work.  An appendix gives technical details of the workings of our method.

\section{Description of the method}
\label{sec:overview}

Suppose we have made some measurements of a networked system.  They may include direct measurements of network structure, such as measurements of the presence or absence or edges, but they may also include other quantities that could have indirect bearing on network structure, such as measurements of node properties or traits.  Our objective is to make the best possible estimate of the structure of the network from these measurements and to do so efficiently and almost automatically, requiring a minimum of technical effort, so that practitioners can focus on making the measurements and interpreting the results.  In this section we give a broad overview of the concepts and essential equations underlying our method.  A complete derivation and accompanying technical discussion can be found in Appendix~\ref{sec:methods}.

In a nutshell, we posit that there exists some underlying network whose structure affects measurements made on the system of interest.  This network is represented by its adjacency matrix~$\bm{A}$, which is initially unknown.  Our goal is to estimate this matrix.  We assume that the observational data depend on the adjacency matrix but in a potentially noisy way: even exact repetition of the same experiment could produce different observations.  This means that, even though the network has a well-defined and unambiguous structure, it will not in general be possible to tell exactly what that structure is from the measurements.  To accommodate this uncertainty, we adopt a Bayesian point of view.  Instead of inferring the exact network structure itself from measurements, we instead infer a probability distribution over possible structures compatible with the observations.

Apart from the data themselves, the only input our method requires of the user is the specification of a model that represents, in general terms, how the data depend on the network structure.  This model can take a variety of different forms: networks and the experiments used to measure them vary widely, so no single model applies in all cases.  Our method allows the user to specify the model that is most appropriate to their situation.  The model may contain parameters (sometimes many of them) but it is not necessary to know the values of these parameters: our method automatically infers the best values from the data.

For the sake of concreteness, we concentrate in this paper primarily on the most common situation encountered in network studies, in which the network is simple, undirected, and unweighted, and the data consist of individual measurements of the presence or absence of edges.  The methods we describe are applicable to other situations as well, but this one covers many cases of interest and will allow us to use a more transparent and explicit notation.  The measurements themselves can be as simple as observed interactions between pairs of nodes, but can also take more complex forms, such as paths across the network, time-series, tags associated with relationships, or any of a variety of other possibilities.  We encode the measurements in an array~$\bm{X}$, whose entry $X_{ij}$ contains all the information we have about the interaction of nodes $i$ and~$j$.

We also make a further crucial assumption about the model, namely that observations of different node pairs are \emph{conditionally independent}, which in this case means that the observations~$X_{ij}$ of the interaction between $i$ and $j$ depend only on the adjacency matrix element~$A_{ij}$ and not on any other elements.  This assumption is not strictly necessary for the method work but, as shown in Appendix~\ref{sec:methods}, it improves the computational efficiency substantially.  And, since it is true of most commonly used models anyway, it is in practice not a significant restriction.  There are exceptions: in certain (``fixed choice'') social network studies, for instance, participants are limited to naming a maximum number of friends or contacts, such as ten.  This means that every time a participant names a contact, contacts with other individuals becomes less likely because the participant has fewer ``slots'' left to fill, and hence contacts are (weakly) negatively correlated.  In this paper we neglect such dependencies and assume that contacts are independent. (However, see Refs.~\cite{newman2018network,newman2018estimating} for a discussion of methods that can handle dependencies.)

With these assumptions, selecting a model boils down to making three basic choices.  The first and most crucial of these is specifying how the pairwise measurements~$X_{ij}$ depend on the underlying network, as represented by the adjacency matrix~$\bm{A}$.  Specifically, for a given pair of nodes~$i,j$ we need to specify the expected distribution of values~$X_{ij}$ for the case where $i$ and $j$ are connected by an edge and for the case where they are not.  We will write these two distributions respectively as
\begin{equation}
\mu_{ij}(1,\theta) = P(X_{ij}|A_{ij}=1,\theta)
\label{eq:mu1}
\end{equation}
and
\begin{equation}
\mu_{ij}(0,\theta) = P(X_{ij}|A_{ij}=0,\theta),
\label{eq:mu0}
\end{equation}
where $\theta$ denotes any parameters of the distribution.  (We will in some cases drop $\theta$ from the notation where it is clearly understood.)  We will refer to Eqs.~\eqref{eq:mu1} and~\eqref{eq:mu0} as the \textit{data model}.

The second modeling choice is the specification of the prior probability ascribed to the edge between $i$ and~$j$.  What is the probability that the edge exists \emph{before} we make any measurements of it?  Do all edges have an equal chance of existing a priori, or are some more likely than others?  Again we assume that different edges are independent, although again this assumption, while computationally convenient, is not strictly necessary.  Mimicking the notation introduced for the data model, we write the prior probability of an edge between $i$ and~$j$ as
\begin{equation}
\nu_{ij}(1,\theta) = P(A_{ij}=1|\theta)
\end{equation}
and the probability of no edge is $\nu_{ij}(0,\theta) = 1 - \nu_{ij}(1,\theta)$.  Again $\theta$ collectively denotes the set of parameters (if any).  We call this second set of probabilities the \emph{network model}.

The third and last modeling choice is the specification of the prior distribution on the parameters~$\theta$.  Our framework does not place any restriction on the possible form of the prior on~$\theta$.  In many cases a simple uniform (maximum entropy) prior works well, but the prior can also be chosen for example to encode specific prior knowledge about the system or to rule out unphysical values of the parameters.

Once these three choices are made, the rest of the procedure is essentially automatic.  Given the model choices and a set of measurements, our method will generate a string of pairs $(\bm{A}_r, \theta_r)$ of networks and parameters compatible with the measurements.  By inspecting these networks and parameters we can determine any other network properties we might care about.  For example, if we want to determine whether $i$ and $j$ are connected by an edge, we can inspect the matrix element~$A_{ij}^{(r)}$ for all~$r$ and compute the fraction of the time that $A_{ij}^{(r)}=1$, which gives us (a Monte Carlo estimate of) the posterior probability of the edge's existence.

More precisely, the sample networks and parameter sets returned by our algorithm are drawn from the joint posterior distribution $P(\bm{A},\theta|\bm{X})$, which allows us to compute an estimate of the expected value of any function~$f(\bm{A},\theta)$ of the network and parameters thus:
\begin{align}
\langle f(\bm{A},\theta) \rangle &= \sum_{\bm{A}} \int f(\bm{A},\theta) P(\theta,\bm{A}|\bm{X})\ d\theta \notag\\
  &\simeq \frac{1}{N}\sum_{r=1}^N f(\bm{A}_r,\theta_r),
\end{align}
where $N$ is the number of samples generated.

Our computer code implementing these calculations is freely available online with accompanying tutorials explaining its use.\footnote{The code can be downloaded from \url{https://github.com/jg-you/noisy-networks-measurements}.}

\section{Examples}
\label{sec:case_studies}

To demonstrate how our method operates in practice, we present two detailed case studies.  The first involves a network of animal interactions.

\subsection{Network of dolphin companionship}

Many animals form lasting social networks, including monkeys, deer, horses, cattle, dolphins, and kangaroos~\cite{whitehead2008analyzing}.  Here we analyze data from Connor~\etal~\cite{connor1992dolphin} of interactions among a small ($n=13$) group of male bottlenose dolphins as they swim in a shallow lagoon.  This is a typical example of an animal observational study that aims to determine social ties indirectly by observing behavior.  Standard techniques of social network analysis would typically be used to transform the observations into association indices that quantify the level of interaction between pairs of individuals~\cite{brask2020animal}.  These indices, however, can can be hard to interpret and their definition relies on somewhat ad hoc assumptions.  Our methods give us a more principled way to infer connections by interpreting the data as noisy measurements of an underlying social network.

\subsubsection{Basic model}
\begin{figure}
\includegraphics[width=\linewidth]{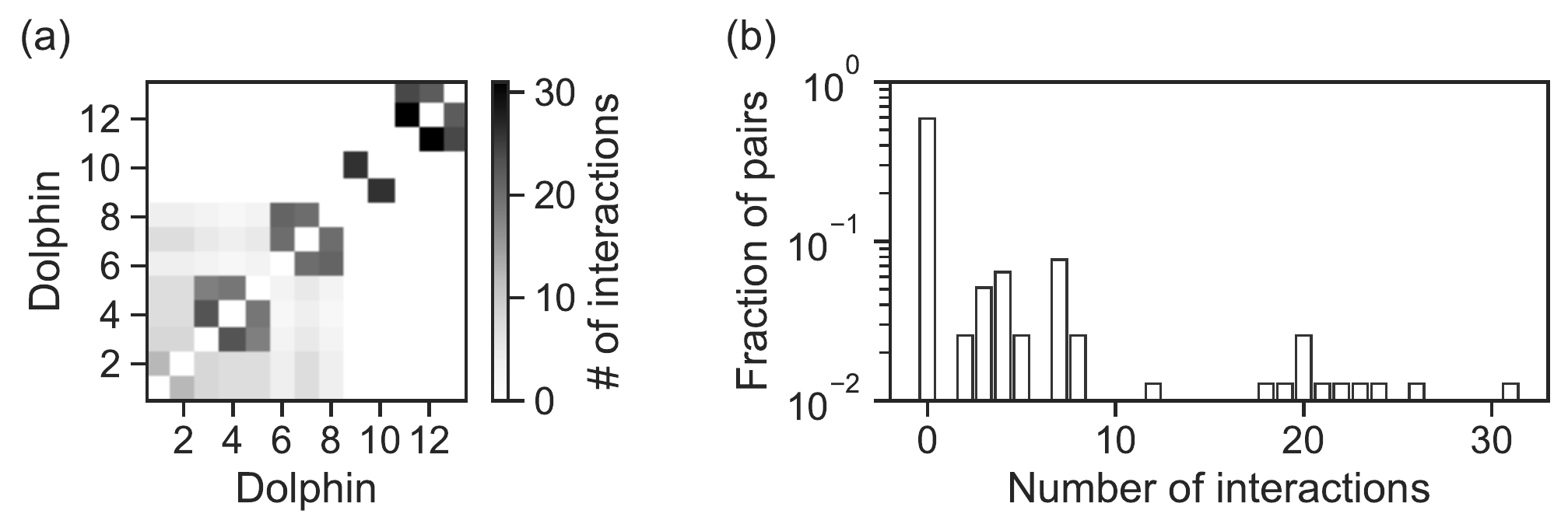}
\caption{Data on interactions among a group of dolphins, from Connor~\etal~\cite{connor1992dolphin}.  Thirteen male dolphins were observed as they swam in a shallow lagoon and tabulations were made of pairs that swam together.  (a)~Observed frequency of interaction for each pair.  (b)~Histogram of frequencies.}
\label{fig:dolphin_data}
\end{figure}

In this particular study the recorded data~$X_{ij}$, shown in Fig.~\ref{fig:dolphin_data}, represent the number of times each pair of dolphins is observed swimming in close proximity.  We can use these data to reconstruct the underlying network as follows.  First, it is reasonable to assume that the number of interactions between two dolphins will depend on whether they have a network connection or not.  But also we expect there to be some randomness in the numbers, both because of circumstances and because of observational error.  We thus model the number of interactions as a Poisson random variable with mean $\lambda_1$ or $\lambda_0$ depending on whether there is or is not a network connection, respectively.  That is,
\begin{subequations}
\label{eq:poisson_model}
\begin{align}
  \mu_{ij}(0,\lambda_0) &= \frac{\lambda_0^{X_{ij}}}{X_{ij}!} e^{-\lambda_0},\\
  \mu_{ij}(1,\lambda_1) &= \frac{\lambda_1^{X_{ij}}}{X_{ij}!} e^{-\lambda_1}.
\end{align}
\end{subequations}
We assume that $\lambda_1>\lambda_0$, i.e.,~that individuals interact more often if they have a network connection than if they do not.

We also need to choose our network model and prior on the model parameters.  Having no prior information about the probabilities of individual network edges, we assume that all edges are a priori equally likely, which implies
\begin{subequations}
\label{eq:ER_prior}
\begin{align}
  \nu_{ij}(0,\rho) &= 1-\rho,\\
  \nu_{ij}(1,\rho) &= \rho.
\end{align}
\end{subequations}
where $\rho$ is the probability of an edge.

For the priors on the parameters we make minimal assumptions.  For~$\rho$, which is a probability, we assume a uniform prior on the interval~$[0,1]$.  For $\lambda_0$ and $\lambda_1$, which have semi-infinite support, we cannot use a uniform prior, so instead we assume a slowly varying probability over a wide range of plausible values.  In keeping with modern Bayesian practice we use a semi-normal distribution with large variance (i.e.,~the right half of a normal distribution centered on zero):
\begin{equation}
P(\lambda_k) \propto e^{-\lambda_k^2/2\sigma^2},
\end{equation}
where $\sigma\gg 1$ is a fixed hyperparameter and $P(\lambda_k)=0$ if $\lambda_k<0$.  In our work we use $\sigma=100$.

\begin{figure}
\begin{minipage}{2.5cm}
\null\vspace{-11.45\baselineskip}
\includegraphics[height=11.\baselineskip]{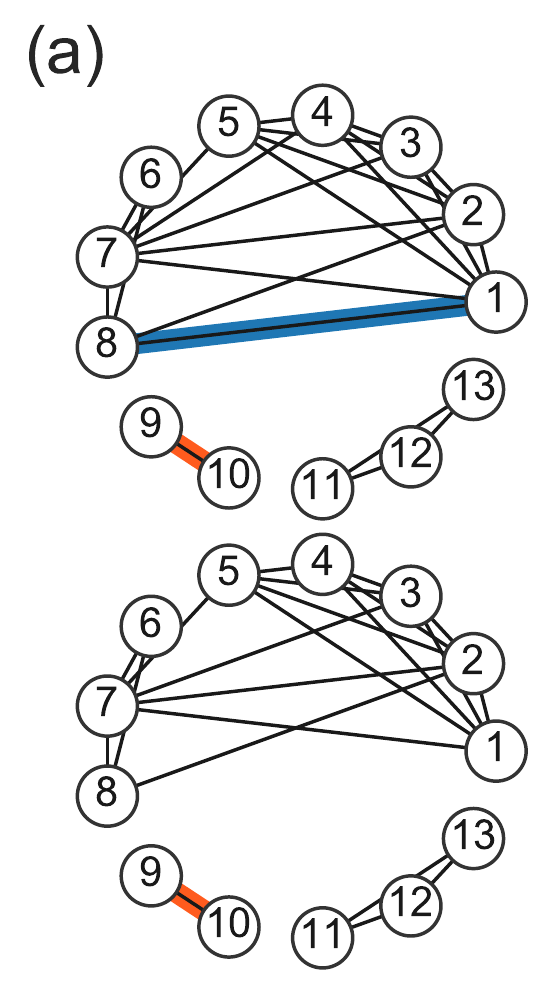}
\end{minipage}
\includegraphics[height=11.5\baselineskip]{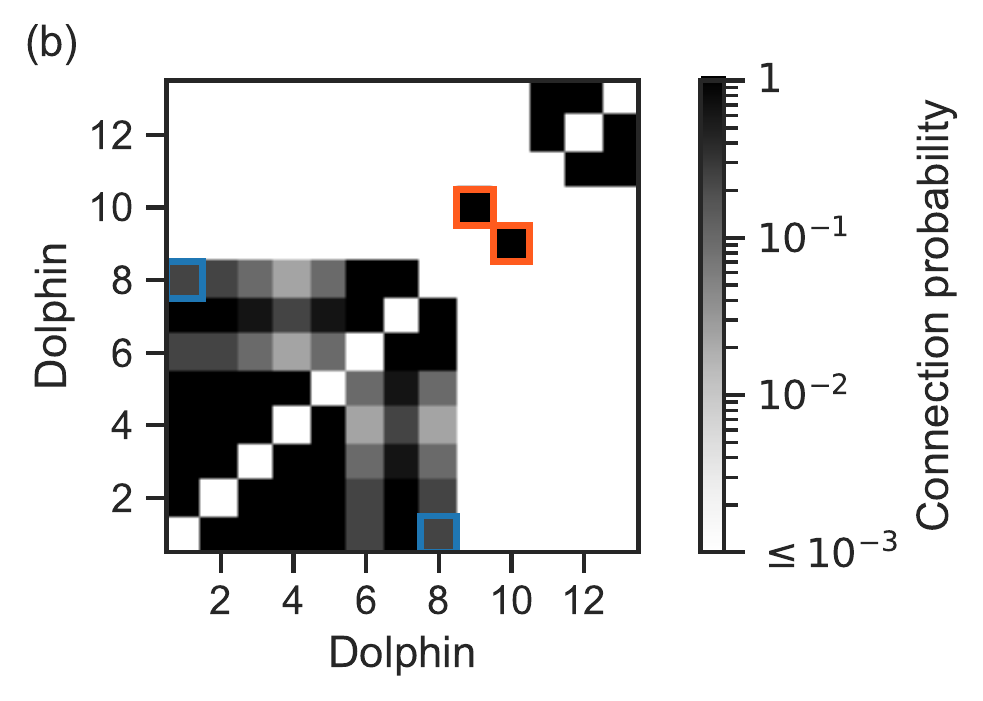}
\caption{Reconstruction of the network of dolphins from the data shown in Fig.~\ref{fig:dolphin_data}.  (a)~Two examples of sampled network structures.  (b)~Matrix of posterior edge probabilities, obtained by averaging over 4000 network samples.  Entries in the matrix corresponding to the edges highlighted in panel~(a) are shown in matching colors.}
\label{fig:dolphin_structure}
\end{figure}

With the model specified, we now run the algorithm described in Appendix~\ref{sec:methods} and obtain a series of samples $(\bm{A}_r,\theta_r)$ from the joint distribution $P(\bm{A},\theta|\bm{X})$, where $\theta$ in this case collectively refers to the parameters $\lambda_0$, $\lambda_1$, and $\rho$, and $\theta_r$ is one realization of these parameters.  Two examples of sampled network structures are shown in Fig.~\ref{fig:dolphin_structure}a, out of thousands generated.  As the figure shows, the two structures are similar but not identical.  This is expected: we should see variability from sample to sample.  When looking over the entire sample set, for example, the edge between nodes 9 and~10, highlighted in orange, appears almost always, whereas the edge between 1 and~8, in blue, appears only rarely.  To capture this variability, we average over samples and compute the posterior probability of every edge as the fraction of samples in which the edge occurs.  The result is shown in matrix form in Fig.~\ref{fig:dolphin_structure}b.  Comparing with Fig.~\ref{fig:dolphin_data}, we see that these probabilities are quite different from the distribution of number of interactions between dolphin pairs.  Moreover, while the numbers of interactions span a wide range of values, the probabilities of edges are clustered around zero and one.  This is good news.  Edge probabilities near zero and one indicate edges about which we are relatively certain, either that they exist or that they do not.  Thus we have turned a data set with considerable variability into a network structure about which we have high confidence.  We discuss this point further below.

In conjunction with these estimates of the network's structure, we also compute estimates of the parameters~$\theta$, by averaging the parameter samples, just as we did with the network samples. For instance, we find that $\lambda_0 = 0.63\pm0.22$ and $\lambda_1 = 14.4\pm1.5$, meaning that dolphin pairs interacted an average of 14.4 times during the experiment if they shared a network connection, but only 0.6 times if they did not.  In other words, the effect of network connections is very pronounced and highly statistically significant.  We also find that $\rho = 0.26\pm0.06$, indicating that the network is quite dense.  (This would be an unusually high value in human social networks, although the network in this example is small, which tends to increase density.)

\subsubsection{Sampling quality and goodness of fit}
While these results are promising, there is further work to be done if we are to be confident in them.  In particular, as is standard with Bayesian calculations, there are two specific things we need to check~\cite{gelman2013bayesian}.  First, we need to be sure that the Monte Carlo algorithm is sampling correctly from the posterior distribution and, second, we need to test whether our proposed model is in fact a good fit to the data.

\begin{figure}
\includegraphics[width=0.47\linewidth]{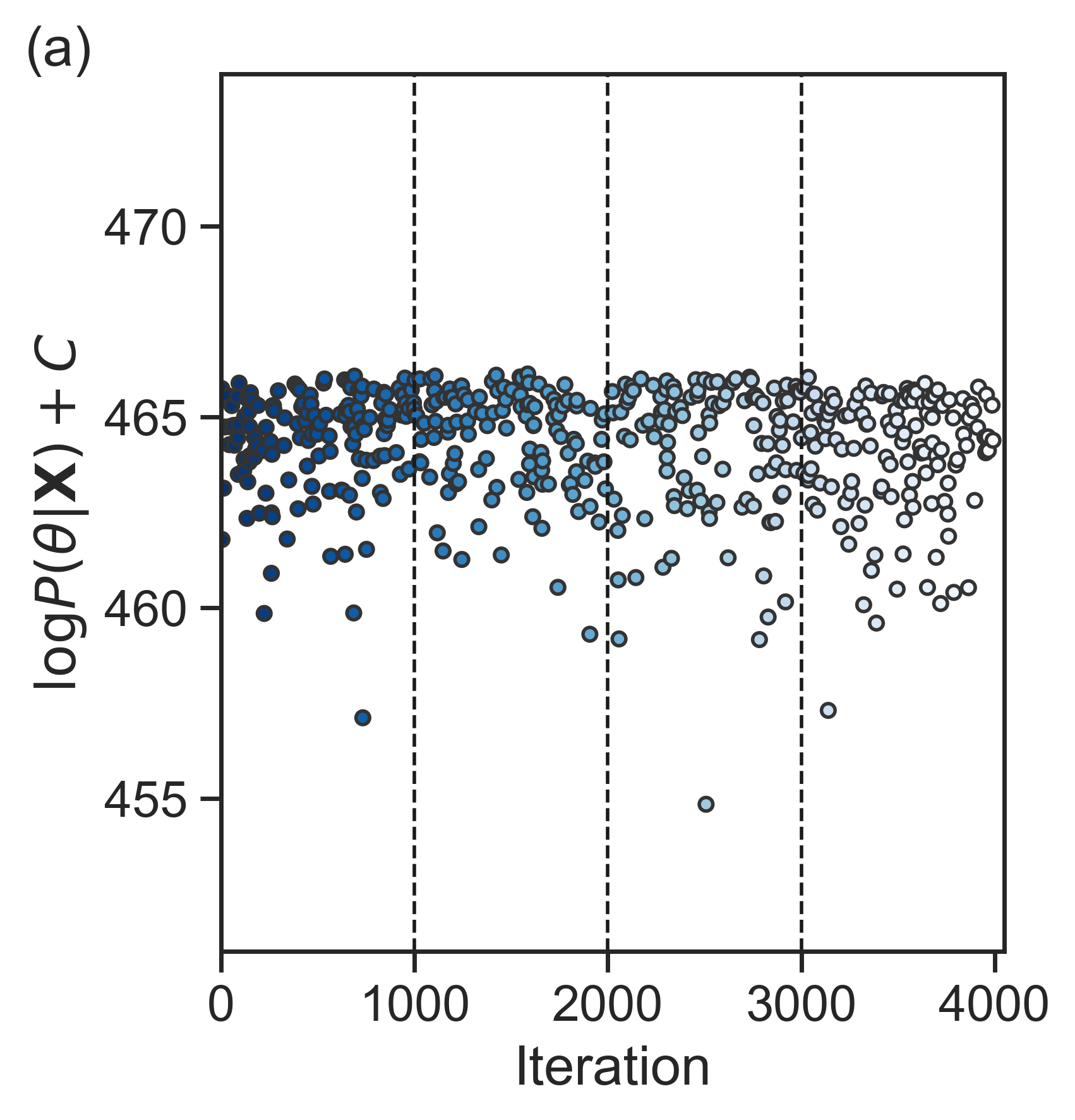}
\includegraphics[width=0.5\linewidth]{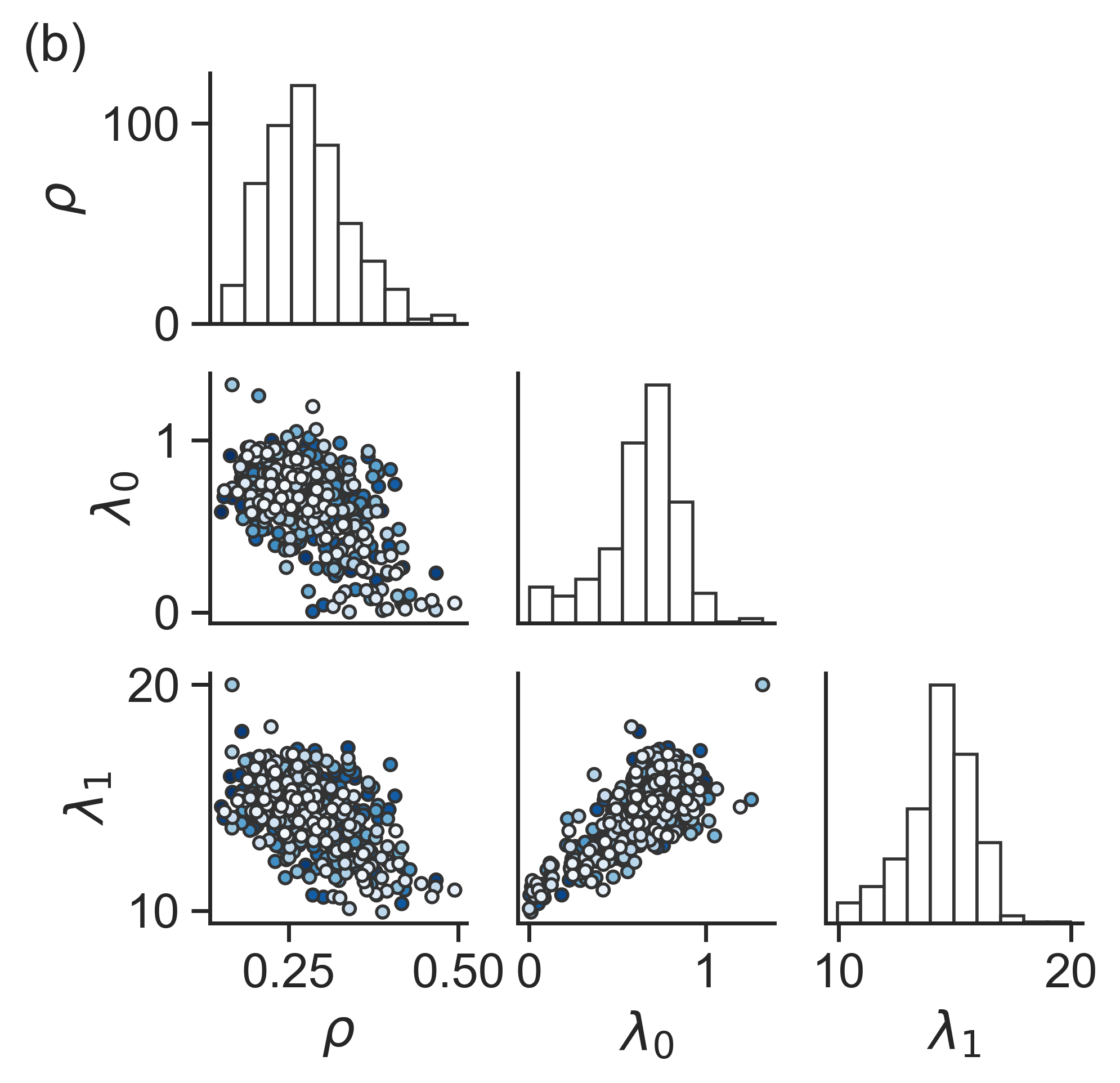}\\ 
\caption{Statistics of samples drawn from the posterior distribution $P(\lambda_0,\lambda_1,\rho|\bm{X})$ for the data shown in Fig.~\ref{fig:dolphin_data} and the measurement model given in Eqs.~\eqref{eq:poisson_model}--\eqref{eq:ER_prior} over four runs with randomly chosen initial conditions.  Only 500 of the 4000 total samples taken are shown for the sake of visual clarity.  Colors indicate the four runs.  (a)~The logarithm of the posterior probability.  Different runs are separated by vertical dotted lines.  (b)~Pair plot showing the relation between sampled values of parameters, as well as the distribution of individual parameters on the diagonal.}
\label{fig:dolphin_diagnosis}
\end{figure}

In Monte Carlo calculations of this kind the posterior probability distribution is typically \emph{rugged}, meaning it has multiple local optima, and the sampling algorithm can as a result get stuck for periods of time in suboptimal portions of the sampling space.  To test for this issue we plot in Fig.~\ref{fig:dolphin_diagnosis}a the log probability of our samples as a function of time over four different runs of the algorithm.  The plot shows that the distribution of values appears roughly consistent within each run and across different runs, which we take as a sign that the samples have not failed in obvious ways.  In Fig.~\ref{fig:dolphin_diagnosis}b we compare how the sampled values for the parameters $\lambda_0$, $\lambda_1$, and~$\rho$ relate to one another.  These plots, colloquially known as pair plots, are conventional in Bayesian analysis.  In addition to showing the distribution of values for the individual parameters on the diagonal (which are more informative than the simple averages reported in the text above), these plots can help verify the quality of the samples and provide some sanity checks.  In our case, the plots reveal that all the samples come from approximately the same region of parameter space regardless of the different initialization of the four runs, which tallies with the uniform sample quality found in Fig.~\ref{fig:dolphin_diagnosis}a and gives further evidence that the algorithm is sampling consistently.

Another function of pair plots is to diagnose problems with the model specification.  Two parameters that cannot be independently determined from the data, for instance, will have strong correlation, and other issues like permutation symmetries will also show up in these plots.  In the case of Fig.~\ref{fig:dolphin_diagnosis}b we find the parameters to be only modestly correlated and there are no major red flags at this stage.

Having confirmed that the samples are plausibly drawn from a correctly specified posterior distribution, the other thing we need to check is whether our model is actually a good fit to the data.  If the model is a poor one, then even the best fit it provides may not actually be \emph{good} fit.  To test the goodness of fit we use two Bayesian tests of the type known as \textit{posterior-predictive assessments}.  (See Appendix~\ref{sec:methods} and Refs.~\cite{gelman2013philosophy,gelman1996posterior} for discussions.)  In these tests we use the data model~$P(\bm{X}|\hat{\bm{A}},\hat{\theta})$ with parameter values $\hat{\bm{A}}$ and $\hat{\theta}$ drawn from our Monte Carlo sample to generate new data~$\tilde{\bm{X}}$, as if we were making measurements on a system that obeyed the fitted model exactly.  Then we compare these new data to the original inputs.  If the model is a good fit, the two should look similar.

\begin{figure}
\includegraphics[width=\linewidth]{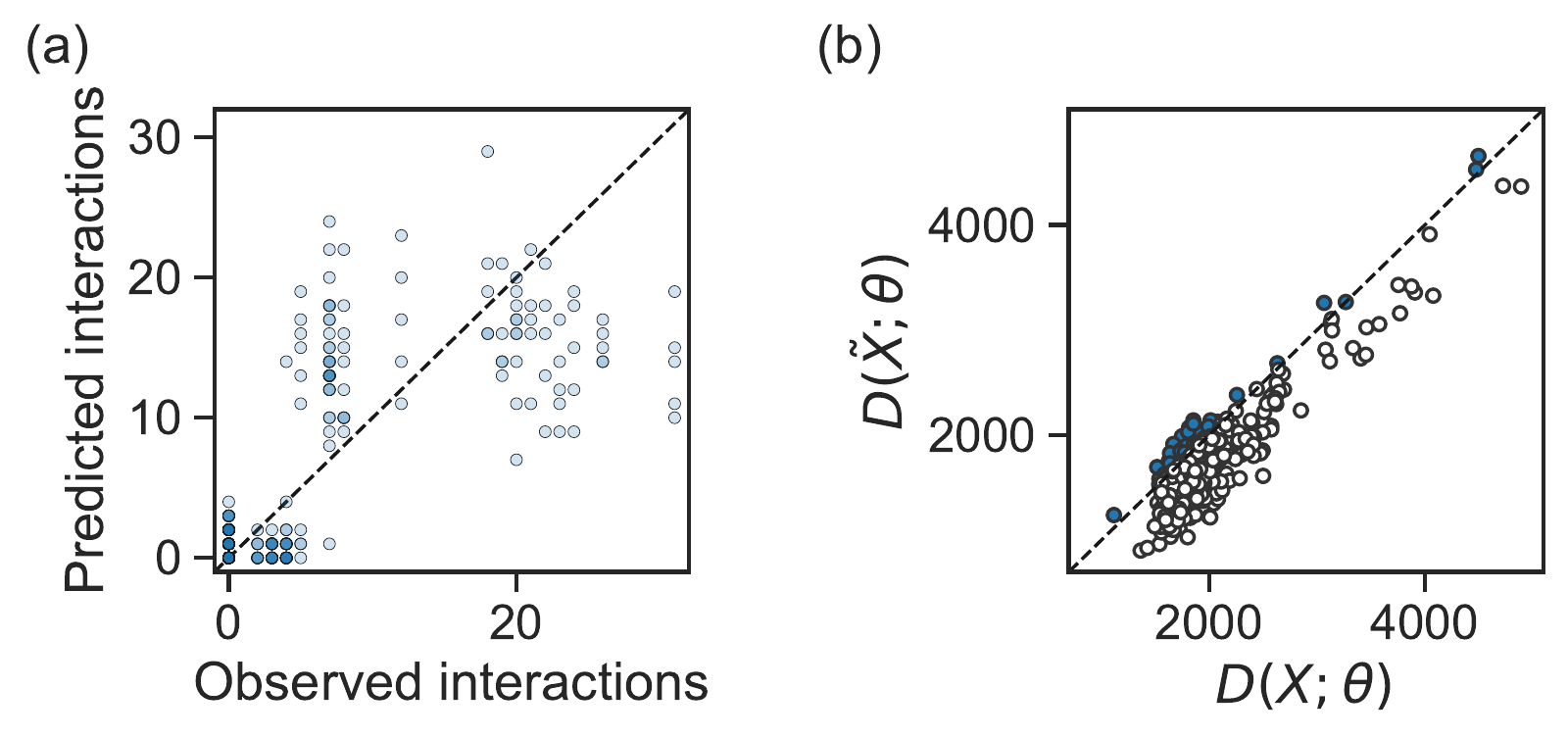}
\caption{Posterior-predictive tests of the model used for the dolphin network.  (a)~Average predicted number of interactions between dolphin pairs as a function of the actual observed number of interactions, for five random data sets generated from the model.  Colors of the data points indicate the numbers of samples and $R^2=0.50$.  (b)~Scatter plot of discrepancy values.  Each point in this plot corresponds to one of 500 sets of parameter values, selected at random from a total of 4000 such sets drawn from the posterior distribution~$P(\lambda_0,\lambda_1,\rho|\bm{X})$ during the calculation.  Sets having higher model--model discrepancy $D(\tilde{X},\lambda_0,\lambda_1,\rho)$ than data--model discrepancy $D(\bm{X},\lambda_0,\lambda_1,\rho)$ are highlighted in blue, above the diagonal.  The fraction of such sets gives us the $p$-value, which in this case is $p=0.136$.}
\label{fig:dolphin_criticism}
\end{figure}

An example of such a comparison is shown in Fig.~\ref{fig:dolphin_criticism}.  Panel~(a) in the figure shows the number of times a pair of dolphins interacts in the simulated data as a function of the number of times they are observed to interact in the original data, in five artificial data sets~$\tilde{\bm{X}}$ generated as above.  If data and model agreed well, most of the points in this scatter plot would concentrate along the diagonal line, but in this case they do not.  This is our first indication that the fit we have found may not be as good as we would like.  We will see shortly how to make the fit much better, but let us proceed for the moment with what we have as an illustration of our goodness-of-fit analysis.

To more accurately quantify the performance of the model we can calculate a \emph{discrepancy} between data sets~\cite{gelman1996posterior}:
\begin{equation}
  D(\bm{X},\theta)  = \sum_{ij} X_{ij} \log \frac{X_{ij}}{\tilde{X}_{ij}(\theta)},
\end{equation}
where $\tilde{X}_{ij}(\theta)$ are the data generated from the model with parameters~$\theta$.  The discrepancy is essentially a Kullback-Leibler divergence between the distribution of real and synthetic data for one sampled value of the parameters.  It functions in this situation as a goodness-of-fit measure: the smaller the discrepancy, the closer the synthetic data are to the input.

The discrepancy is not very informative by itself, however, since it is not clear what kind of values we should expect to see.  For example, even if the model is a perfect fit, we should not expect the discrepancy to be zero, since the randomness of both the data and the model mean that they are unlikely to agree exactly.  To obtain a point of comparison, therefore, we compute discrepancies between a large number of pairs of simulated data sets drawn from realizations of the model with the same parameter values used for the observed data.  If the model were correct, so that simulated and observed data have the same distribution, then these values would tell us the typical magnitude we should expect the discrepancy to have.  If the values are mostly smaller than the observed discrepancy then it indicates the model is unlikely to be correct; if they are larger then the model is not ruled out.  The fraction of generated discrepancy values that are larger than the observed discrepancy gives us the $p$-value for the model, i.e.,~the probability that the observed discrepancy would have been generated if the model were correct.

Note that the use of the $p$-value in this situation is different from the way it is used in traditional frequentist statistics, where it represents the probability of getting a particular observed value if a null hypothesis were true.  In the traditional scenario a small $p$-value leads us to reject the null hypothesis, and, since this is often the goal of an experiment, small $p$-values are ``good.''  In the present case, the $p$-value is applied to the model we are fitting (there is no null/alternative hypothesis) and it just counts the fraction of artificial data sets for which we find discrepancies more extreme than the value found when fitting the model to the true data.  So in this situation small $p$-values are ``bad.''

Figure~\ref{fig:dolphin_criticism}b shows values of the discrepancy for both artificial and real data, for 500 sampled values of the model parameters.  Instances where the artificial discrepancy is greater than the observed one appear above the diagonal in the plot and the fraction of such points tells us the $p$-value.  In this case we find $p=0.136$.  While it is not appropriate to apply arbitrary cutoffs to this (or any) $p$-value~\cite{gelman1996posterior}, the value is not as high as we would like it to be, and though we cannot completely rule out the model the evidence suggests that the fit is not ideal.

\subsubsection{Improved model}
What can we do to improve the fit?  The standard approach is to adopt a more elaborate model that is capable of representing a wider range of data distributions.  In the model we have used so far connections are binary: dolphin pairs either have a connection or they don't.  We can create a more nuanced model by allowing three levels of connection, corresponding to no tie, a weak tie, or a strong tie.  Denoting the three levels by adjacency matrix elements $A_{ij} = 0$, 1, and~2, we introduce a new distribution for $X_{ij}$ when $A_{ij}=2$ which is Poisson as before but with mean~$\lambda_2$:
\begin{equation}
  \mu_{ij}(2,\lambda_2) = \frac{\lambda_2^{X_{ij}}}{X_{ij}!}e^{-\lambda_2}
\end{equation}
where $\lambda_2\ge\lambda_1\ge\lambda_0$, and the prior of Eq.~\eqref{eq:ER_prior} becomes
\begin{subequations}
\label{eqs:multi_ER_prior}
\begin{align}
  \nu_{ij}(0,{\rho}) &= \rho_0 = 1 - \rho_1 - \rho_2, \\
  \nu_{ij}(1,{\rho}) &= \rho_1,\\
  \nu_{ij}(2,{\rho}) &= \rho_2.
\end{align}
\end{subequations}
The fitting and model verification procedures follow the same lines as previously.

\begin{figure}
\includegraphics[width=0.9\linewidth]{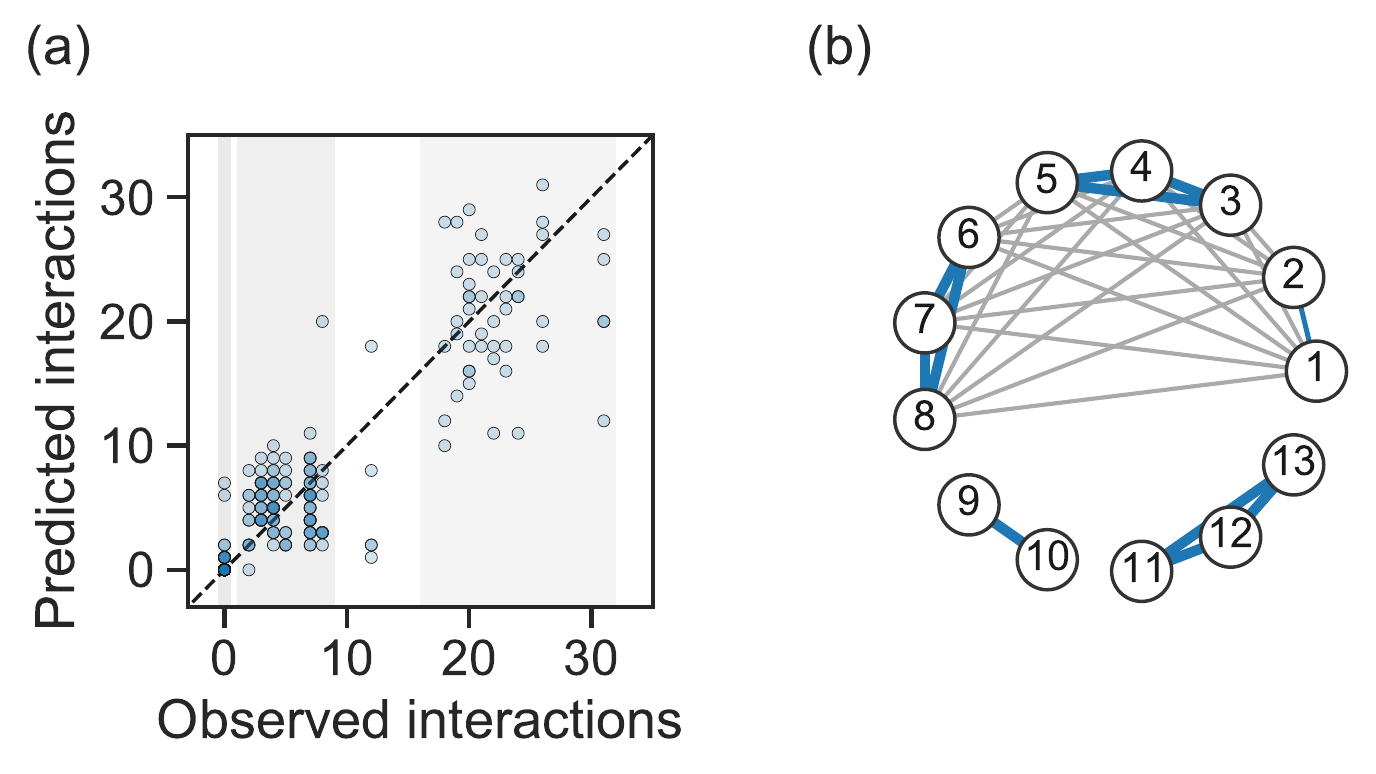}
\caption{Network inference with multiple edge types for the data shown in Fig.~\ref{fig:dolphin_data} and the measurement model given in Eqs.~\eqref{eq:poisson_model}--\eqref{eqs:multi_ER_prior}.  The estimated mean numbers of interactions are $\lambda_0\simeq 0.02$ when there is no edge between a pair of dolphins, $\lambda_1=5.13$ when the pair shares a weak tie, and $\lambda_2=21.97$ when they share a strong one.  The corresponding prior probabilities of edge types are $\rho_0=0.58$, $\rho_1=0.28$, and $\rho_2=0.14$.  (a)~Posterior predicted number of interactions between dolphins as a function of observed number for five random data sets generated from the fitted model ($R^2=0.82$).  The highlighted regions correspond, from left to right, to dolphins having no tie, a weak tie, and a strong tie.  (b)~The inferred structure of the network, with weak ties represented by thin gray edges and strong ties by thicker blue edges.  (Nodes~1 and 2 are connected by a thin blue line to reflect the fact that the calculation is ambiguous about the type of this tie.)}
\label{fig:dolphin_improved}
\end{figure}

This modified model now fits the data significantly better, as shown in Fig.~\ref{fig:dolphin_improved}a.  It divides the observed numbers of pair interactions into three clear groups centered around values of about 0, 5, and~25, and the $p$-value is now a very respectable~$0.722$, indicating that there is no statistical basis to reject this model at all: the model truly captures the structure of the observed behavior.  Indeed a $p$-value significantly larger than this could be a sign of problems, indicating overfitting.  Unless the distribution of the discrepancy is strongly skewed, the expected $p$-value will normally be around 0.5 when the model is a perfect fit.

Having found a model that fits the data well, we examine the inferred network structure, which is shown in Fig.~\ref{fig:dolphin_improved}b.  The network has three disconnected subgroups of dolphins, two comprised of strong connections only and one, the largest of the three, having a mix of strong and weak connections.  The posterior probabilities on all of the interaction types are close to one, indicating high confidence in the structure of the network.  For instance, the model predicts that nodes 8 and 9 are not connected with probability $0.99(6)$, that nodes 7 and 8 share a strong connection with probability $0.99(9)$, and that nodes 1 and 8 share a weak connection with probability~$0.99(9)$.  There is just one pair of nodes (1~and~2) whose connection is hard to classify.  The model indicates that the tie between these nodes is either weak or strong, with probabilities $0.51(4)$ and $0.48(5)$, respectively.  This pair of dolphins was observed swimming together 12 times, a number that falls between the weak and strong domains in the fitted model (see Fig.~\ref{fig:dolphin_improved}a).

\subsection{Friendship network of school students}

For our second example, we revisit a network analyzed previously using different methods in Refs.~\cite{newman2018network,newman2018estimating}, a network of friendships between high-school students taken from the US National Longitudinal Study of Adolescent to Adult Health (the ``AddHealth'' study).  Although the method proposed in this paper is mathematically more complex than that of~\cite{newman2018network,newman2018estimating}, it is arguably easier to apply since our analysis pipeline performs most of the calculation automatically.  The most demanding step is the formulation of the model, but once we have a plausible model the rest of the process is straightforward and mechanical.

The AddHealth study is a large study of networks of social contact among students in schools across the United States.  Students in participating schools were asked to identify their friends and the basic unit of resulting data is a friendship nomination: one student says they are friends with another.  Our data matrix~$\bm{X}$ in this case is thus a non-symmetric one: $X_{ij}=1$ if $i$ names $j$ as a friend and 0 otherwise.  There is no guarantee that $j$ will also name~$i$ and in fact there are many instances in which reported friendships only run in one direction.  If we assume that friendship is fundamentally a bidirectional interaction then this lack of symmetry indicates that the data are necessarily unreliable.

As is done in Refs.~\cite{butts2003network,newman2018network,newman2018estimating}, we fit the data using a model in which each student~$i$ has an individual true-positive rate~$\alpha_i$ and false-positive rate~$\beta_i$.  The true-positive rate is the probability that $i$ names as a friend another student who is in fact a friend, as determined by the adjacency matrix.  The false-positive rate is the probability of naming someone who is not actually a friend.  We explicitly allow for different true- and false-positive rates for different individuals, since it is widely accepted that survey respondents vary in the accuracy of their responses.

In the notation of this paper the equations for the model are:
\begin{subequations}
\begin{align}
  \mu_{ij}(1,\alpha_i,\alpha_j) &= \alpha_i^{X_{ij}}(1 - \alpha_i)^{1 - X_{ij}} \alpha_j^{X_{ji}}(1 - \alpha_j)^{1 - X_{ji}}, \\
  \mu_{ij}(0,\beta_i,\beta_j) &= \beta_i^{X_{ij}}(1 - \beta_i)^{1 - X_{ij}}  \beta_j^{X_{ji}}(1 - \beta_j)^{1 - X_{ji}}.
\end{align}
\end{subequations}
For instance, supposing that $i$ and $j$ truly are friends, the probability of $i$ saying that they are ($X_{ij}=1$) while $j$ says they are not ($X_{ji}=0$) is $\mu_{ij}(1) = \alpha_i ( 1 - \alpha_j)$.  Conversely, if they are not in fact friends then we instead get $\mu_{ij}(0) = \beta_i(1 - \beta_j)$.

For the priors we again make the assumption of Eq.~\eqref{eq:ER_prior} that all edges are a priori equally likely, and assume a uniform prior on the edge probability~$\rho$ and a uniform distribution over all values of $\alpha_i$ and $\beta_i$ that satisfy $\beta_i<\frac{1}{2}<\alpha_i$.  (One could simply assume a uniform prior on both $\alpha_i$ and $\beta_i$ in the range $[0,1]$ but this leaves some ambiguity in the model because of the inherent symmetric between edges and non-edges: if we exchange the values of all $\alpha_i$ and all $\beta_i$ and set $\rho$ to $1-\rho$ the model remains the same.  By making the reasonable assumption that $\alpha_i>\beta_i$ we break this symmetry.  The assumption that $\beta_i<\tfrac12<\alpha_i$ is not strictly necessary, but turns out to be helpful for narrowing down the parameter space and hence improving the speed and convergence of the calculation~\cite{butts2003network}.)

\begin{figure}
\centering
\includegraphics[width=0.6\linewidth]{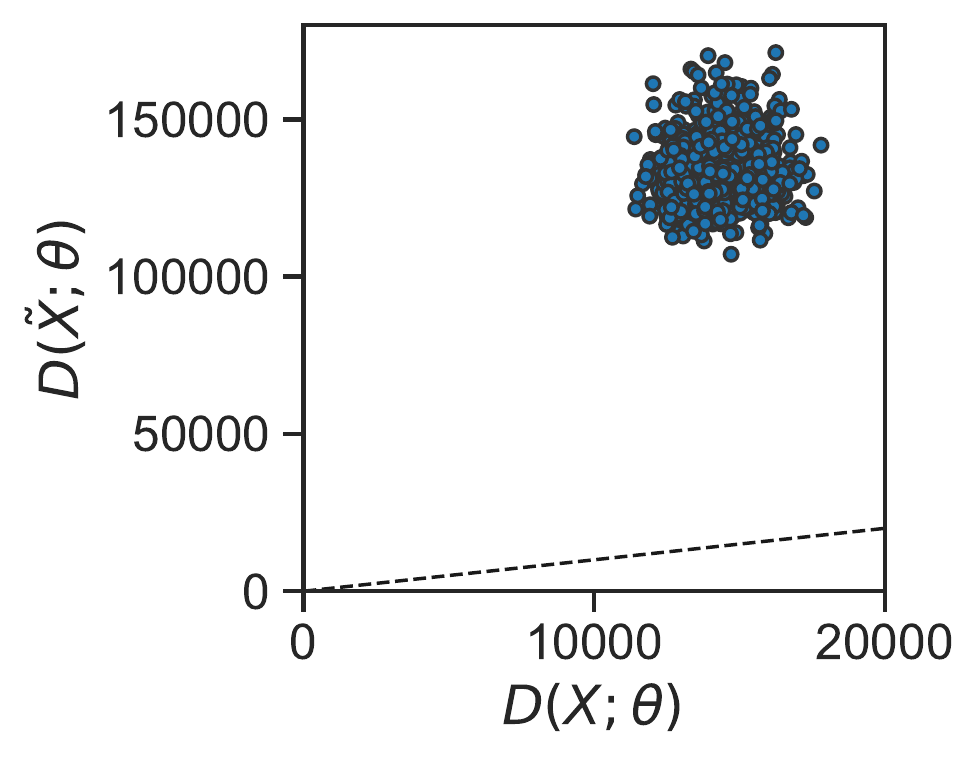}
\caption{Goodness of fit testing for the AddHealth model.  Each point in this plot corresponds to a random parameter set drawn from the posterior distribution~$P(\alpha,\beta,\rho|\bm{X})$.  Samples associated with a higher model--model discrepancy $D(\tilde{X},\theta)$ than data--model discrepancy $D(\bm{X},\theta)$ appear above the dotted line, indicating a good fit between data and model.}
\label{fig:addhealth_criticism}
\end{figure}

\begin{figure}
\centering
\includegraphics[width=0.85\linewidth]{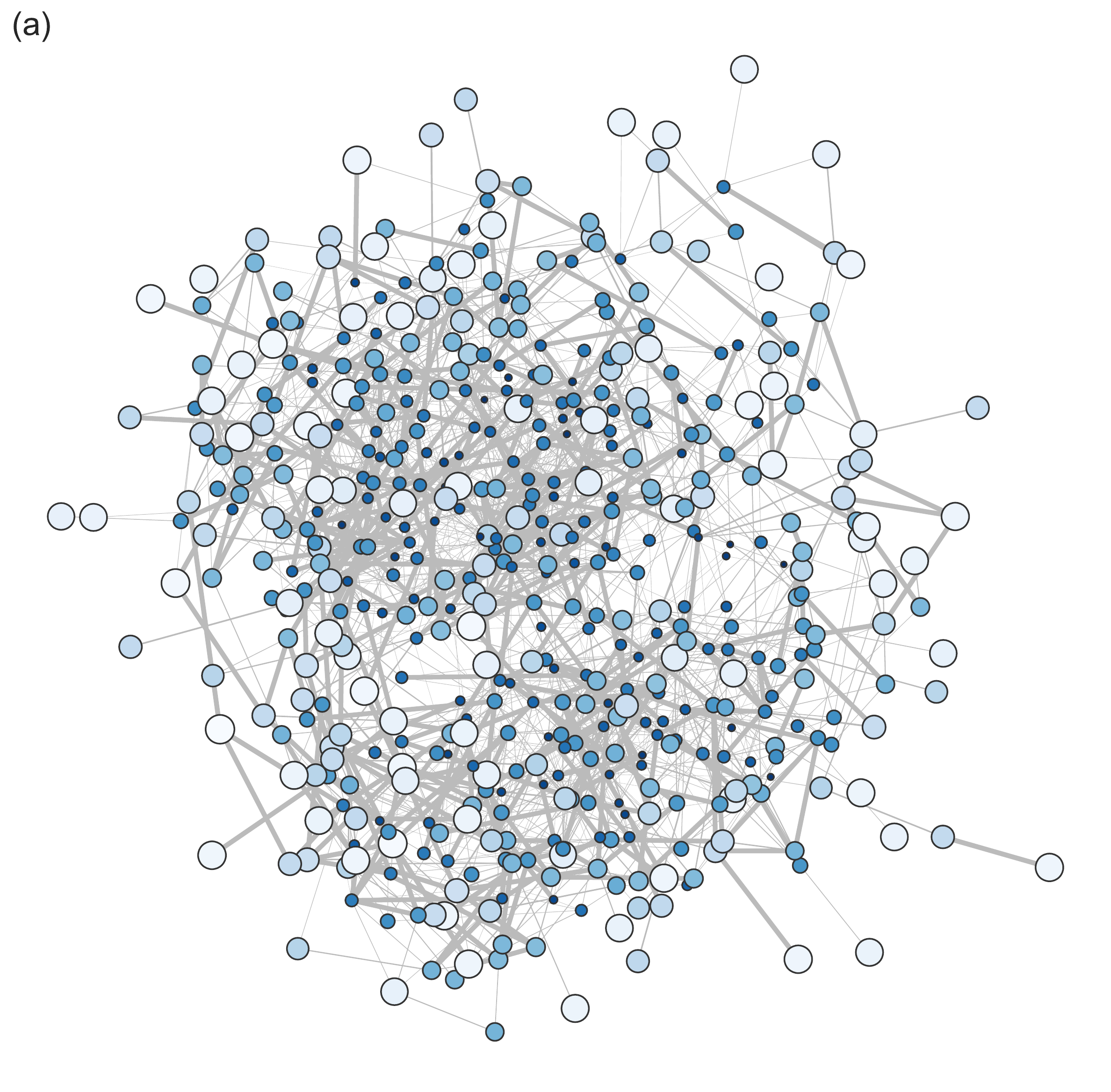}\\
\includegraphics[width=0.8\linewidth]{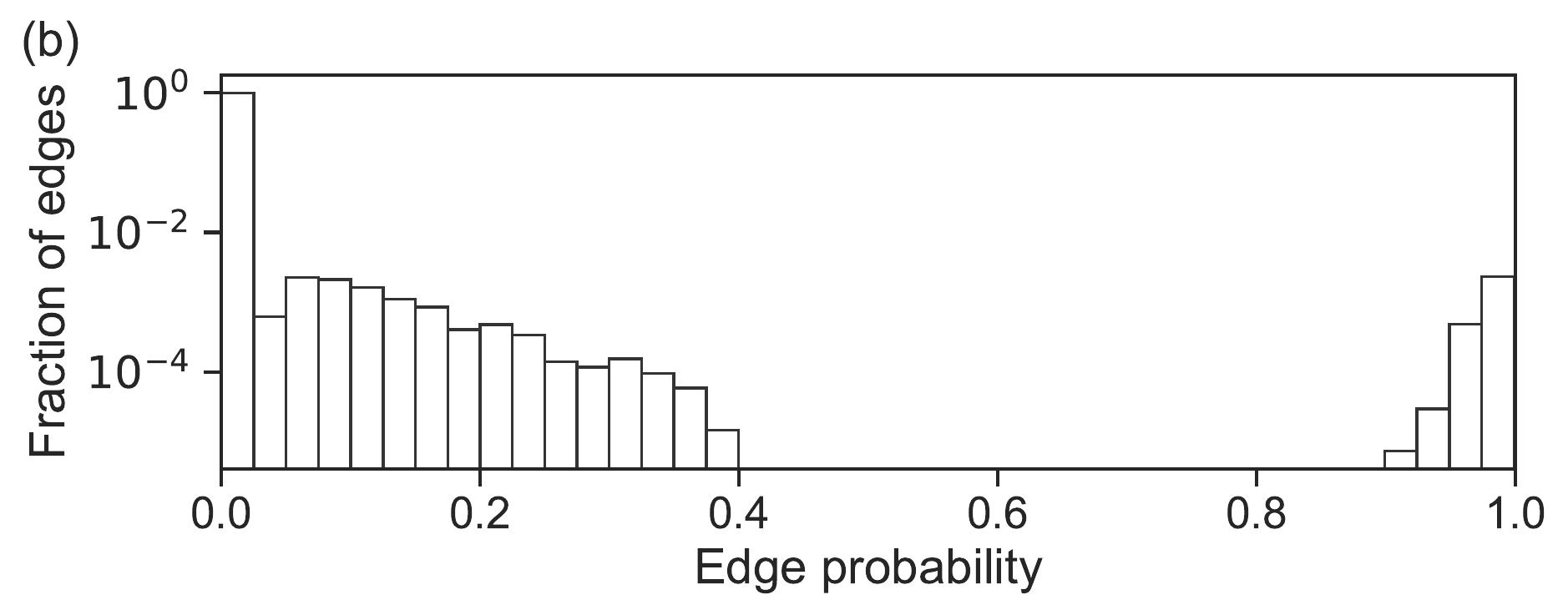}
\includegraphics[width=0.8\linewidth]{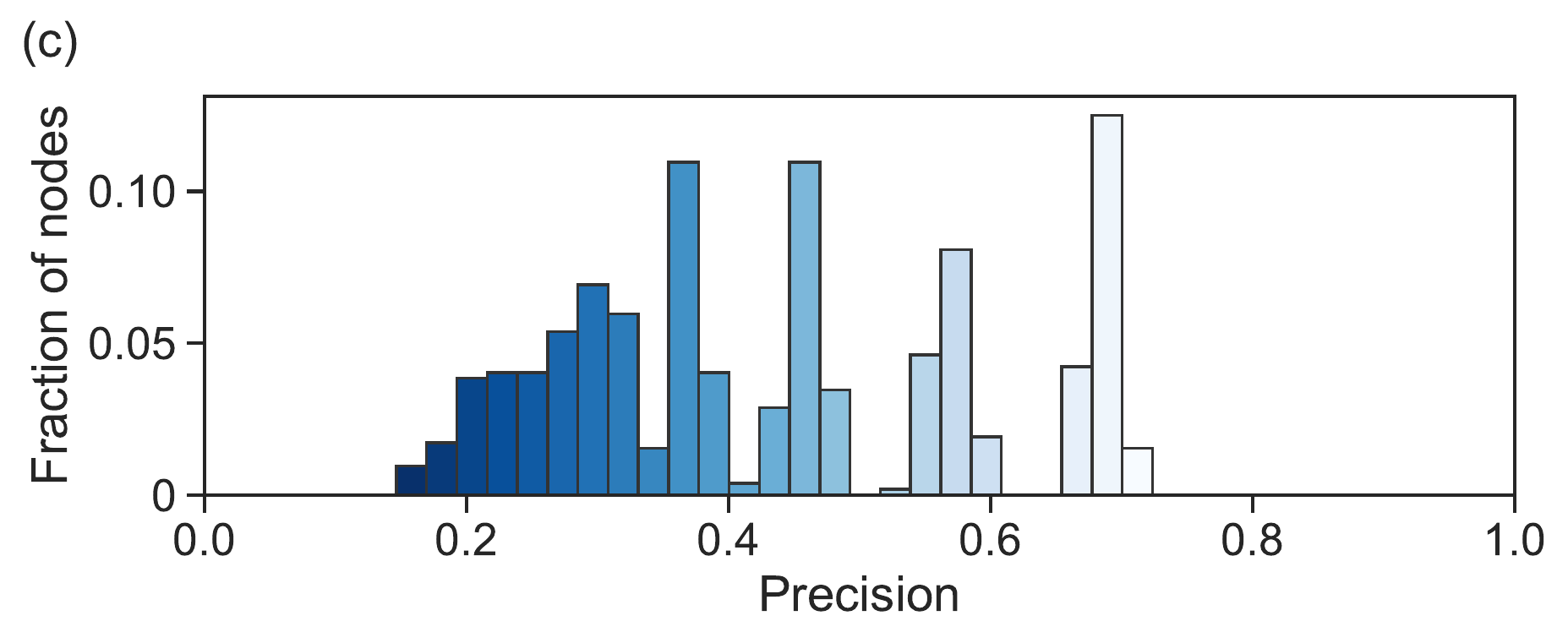}
\vspace{\baselineskip}
\caption{Inference of a school friendship network from noisy data.  (a)~The 521 nodes in this figure represent the students at a single school in the AddHealth study and inferred friendships are shown as edges whose thickness indicates the estimated probability that they exist.  The size and color of the nodes indicates the estimated precision of friendship reports by the corresponding individual, i.e.,~the fraction of their reported friendships that are inferred to actually exist.  Darker shades indicated less precise individuals and correspond to the shades in the histogram in panel~(c).  The average values of the parameters of the model are $\langle \alpha\rangle \simeq 0.7605$, $\langle \beta\rangle \simeq 0.0065$, and $\langle \rho \rangle \simeq 0.004$.  (b)~The distribution of the probability of existence of edges.  Many values are close to zero or one, indicating confidence that the corresponding edge does or does not exist, although a significant number fall at intermediate values.  (c)~The distribution of estimated precision values for participants.}
\label{fig:addhealth_fit}
\end{figure}

Figures~\ref{fig:addhealth_criticism} and~\ref{fig:addhealth_fit} show the results of fitting this model to the data for a single school from the AddHealth data set.  We use one of the smaller schools as our example, with 521 students who completed a survey and 2182 declared ties, primarily in order to make visualization of the results easier.  We find that the Monte Carlo algorithm converges well and gives samples that appear to accurately characterize the posterior distribution.  Figure~\ref{fig:addhealth_criticism} shows discrepancy values in a manner analogous to Fig.~\ref{fig:dolphin_diagnosis}b for the dolphin network and all values are well above the diagonal, indicating a good fit to the data.

The inferred network structure is shown in Fig.~\ref{fig:addhealth_fit}a.  By contrast with the dolphin network example, the posterior probabilities of edges now vary more widely, as represented by the thickness of the edges in the figure.  Figure~\ref{fig:addhealth_fit}b shows the distribution of edge probabilities as a histogram and many probabilities are again close to either 1 or~0, indicating a high degree of certainty that these edges either exist or do not, but there are also a significant number of edges with intermediate probabilities, edges about which we are less certain.

The fit also returns values of the true- and false-positive rates $\alpha_i$ and $\beta_i$ for each node, which allow us to make quantitative statements about how accurately each individual reports his or her friendships.  The average value of~$\alpha_i$ over all individuals and all samples is~0.76, meaning that an estimated 24\% of friendships are going unreported.  The average false-positive rate is 0.0065, which sounds small but this result is somewhat misleading.  The network is very sparse, meaning that almost all edges that could exist do not.  It takes only a small fraction of false-positives among these many non-edges to generate a significant number of errors.  Arguably a more informative measure of false positives is the \emph{precision}, which is the fraction of reported friendships that are actually present and is given in this case by $\rho \alpha_i /[ \rho \alpha_i + (1 - \rho) \beta_i]$~\cite{newman2018estimating}.  The distribution of values of the precision is shown in Fig.~\ref{fig:addhealth_fit}c and ranges from a little under 0.2 to a little over 0.75, indicating that in fact a significant fraction of reported friendships---anywhere from 25\% to 80\%---are false positives.

These results are largely in agreement with previous work~\cite{newman2018estimating}, although there are modest differences in estimated parameter values and network structure, due to the different methodology.  We would argue that the fully Bayesian methodology employed here is more correct in that it accounts for intrinsic uncertainty in the parameter values.  The methodology of~\cite{newman2018estimating}, which makes use of an expectation-maximization (EM) algorithm, might be described as ``semi-Bayesian,'' computing a full posterior distribution over the network structure but relying on point estimates of the parameters.  Because the model used here is a large one, having $\Ord(n)$ parameters, we expect there to be significant uncertainty in the parameter values, which is captured by our Bayesian sampling method.  That said, in practice the two methods do lead to qualitatively consistent conclusions.  The key benefits of the current approach in this case are that it is simpler to implement using standard software, is formally more correct, and incorporates a natural means for checking the goodness of fit.

One potential issue with the results is the fact that the discrepancy values in Fig.~\ref{fig:addhealth_criticism} are all well above the dotted line, indicating close fits of the model to the data and a $p$-value near~1.  A $p$-value this large can be a warning sign for overfitting, which is a possibility given the large number of parameters in the model.  Such an issue could not be diagnosed with the methods previously used in Refs.~\cite{newman2018network,newman2018estimating}, but our approach makes this possible.  One could address the problem by changing the model, say by using a more complex model in which instead of fitting the true- and false-positive parameters we instead draw them from a hyperprior distribution, such as a beta distribution, with an associated (small) set of hyperparameters that are fit using Monte Carlo.  This approach can reduce the chances of overfitting and would be a good direction for future work.

\section{Conclusions}
\label{sec:concs}
In this paper we have introduced a general Bayesian framework for reconstructing networks from observational data in the case where the data are error prone, even when the magnitude of the errors is unknown.  Our methods work by fitting a suitable model of the measurement process to the data and there is a large class of models that is both expressive enough to represent real data sets accurately and yet simple enough to allow for easy and automatic statistical inference.  The output of the fitting process is a complete Bayesian posterior distribution over possible network structures and possible values of model parameters.  We have demonstrated our methods with two case studies showing how to formulate suitable models, fit them, assess goodness of fit, and infer reliable estimates of network structure.

With this work, we hope to promote the adoption of more rigorous methods for handling measurement error in network data in a principled manner.  The methods we propose not only achieve this but do so in a manner that is straightforward and requires a minimum of technical expertise on the part of the user.  Practitioners can use the framework we propose to apply appropriate, application-specific models to their data and obtain estimates of network structure in a matter of minutes.

\section*{Acknowledgments}
We thank Alec Kirkley for helpful discussions.  This work was funded in part by the James S. McDonnell Foundation (JGY) and the US National Science Foundation under grant DMS--1710848 (MEJN).

\appendix
\section{Methods}
\label{sec:methods}

In this appendix we describe the mathematical and statistical foundations of our method in detail.

\subsection{Generative models of measurement}
Consider an experimental setting in which we have measurements~$\bm{X}$ of a network's structure.  The measurements could be as simple as a number of observed interactions between pairs of nodes, but could also incorporate time-series, vector measurements, etc.  In general these measurements do not tell us the exact structure of the network, but instead give us indirect and potentially noisy information.  Our goal is to make the best estimate we can of the true network structure given the measurements.

In the general framework we consider here, two nodes $i$ and $j$ can share connections of various types.  In the simplest case there are just two types: nodes can be either connected by an edge (type~1) or not (type~0).  In a more complex three-type case the connection could be absent (type~0), weak (type~1), or strong (type~2), and so on.  For a network of $n$ nodes we encode these connections by an $n\times n$ adjacency matrix~$\bm{A}$ where the matrix element~$A_{ij}$ records the type of connection between nodes $i$ and~$j$.  We can also represent directed networks using an asymmetric adjacency matrix with $A_{ij}$ being the type of the directed connection from $j$ to~$i$ and $A_{ji}$ being the type from $i$ to~$j$.

Our approach rests on the hypothesis that the matrix~$\bm{X}$ of pairwise measurements is dependent, in a probabilistic fashion, on the adjacency matrix~$\bm{A}$.  Both $\bm{A}$ and $\bm{X}$ can be either symmetric (for undirected networks) or asymmetric (for directed ones) and they need not be of the same type.  In friendship networks, for example, the symmetric relationship of being friends is commonly probed using asymmetric measurements (person~$i$ says they are friends with person~$j$).

It is this dependence between network and measurement that we exploit to estimate $\bm{A}$ from~$\bm{X}$.  We formalize the relation using a generative model that specifies the probability~$P(\bm{X}|\bm{A},\theta)$ of making the measurements given the network, plus optionally some additional parameters represented collectively by~$\theta$.  Then, applying Bayes' rule, we can write the probability of the unknown quantities~$\bm{A}$ and~$\theta$ given the measurements as
\begin{equation}
\label{eq:bayes}
P(\bm{A},\theta|\bm{X}) = \frac{P(\bm{X}|\bm{A},\theta)P(\bm{A}|\theta)P(\theta)}{P(\bm{X})}.
\end{equation}
Our goal is to use this equation to infer the network structure $\bm{A}$ from the measurements $\bm{X}$ and to quantify the errors we might make in doing so.

\subsection{A flexible class of models}
\label{sec:model_class}

To further simplify the discussion and improve the efficiency of the numerical calculations we make some additional assumptions about the model, while keeping the approach as broad as possible to allow users to easily adapt it to various types of data and experimental settings.

Of the four probabilities that appear on the right-hand side of Eq.~\eqref{eq:bayes} one of them $P(\bm{X})$ is a constant (since it depends only on~$\bm{X}$ which is fixed by the experiment) and hence will play no part in our calculations.  The others must be specified to define our model.  We refer to these three probabilities as the data model $P(\bm{X}|\bm{A},\theta)$, the network model $P(\bm{A}|\theta)$, and the prior on the parameters~$P(\theta)$.  Let us consider each of these in turn.

\subsubsection{Data model}
The data model $P(\bm{X}|\bm{A},\theta)$ specifies the probability of making a particular set of measurements~$\bm{X}$ given the network and the model parameters.  In specifying this probability we will make two key assumptions.  First, we assume that the measurement~$X_{ij}$ is only influenced by the corresponding element~$A_{ij}$ of the adjacency matrix and not by any other elements.  Second, we assume that, conditioned on the network structure~$\bm{A}$ and parameters~$\theta$, the measurements~$X_{ij}$ for different node pairs are independent.
Thus, for instance,
\begin{equation*}
  P(X_{ij},X_{kl}|\bm{A},\theta) = P(X_{ij}|A_{ij},\theta) P(X_{kl}|A_{kl},\theta).
\end{equation*}

The notation here is a bit unwieldy, so for clarity we introduce the notation $\mu_{ij}(A_{ij},\theta)$ to denote the probability $P(X_{ij}|A_{ij},\theta)$ of making the measurement~$X_{ij}$ given the type~$A_{ij}$ of the connection between nodes $i$ and~$j$ and given the parameter values~$\theta$.  (Where the meaning is clear we may drop the explicit dependence on~$\theta$ to simplify our expressions.)  With this notation and our assumption of conditional independence, the probability~$P(\bm{X}|\bm{A},\theta)$ for the data model is simply
\begin{equation}
\label{eq:data_model}
P(\bm{X}|\bm{A},\theta) = \prod_{(i, j)} \mu_{ij}(A_{ij},\theta).
\end{equation}
The product $\prod_{(i,j)}$ is taken over all unordered pairs of nodes when the network is undirected and over all ordered pairs when it is directed.

Table~\ref{table:data_models} gives a selection of possible forms for the data model for networks with only a single edge type.  Generalization to multiple edge types is straightforward.  (See also Ref.~\cite{newman2018estimating} for a discussion of a range of models.)

\subsubsection{Network model}
The network model $P(\bm{A}|\theta)$ can be thought of as our prior expectation of what the network should look like, before we make the measurements.  By analogy with the factorized form of the data model in Eq.~\eqref{eq:data_model}, we consider network models with the factorized form
\begin{equation}
  \label{eq:network_model}
  P(\bm{A}|\theta) = \prod_{(i, j)} \nu_{ij}(A_{ij},\theta),
\end{equation}
where we define $\nu_{ij}(A_{ij},\theta)$ in a similar manner to~$\mu_{ij}(A_{ij},\theta)$, as the prior probability $P(A_{ij}|\theta)$ that nodes $i$ and $j$ share a connection of type~$A_{ij}$, given the parameters~$\theta$.  Many standard network models can be written in this form, including the Erd\H{o}s--R\'enyi random graph, the configuration model, and the stochastic block model.  Some examples of network models are given in Table~\ref{table:network_models} and Ref.~\cite{newman2018estimating}.

\subsubsection{Prior on the parameters}
The third component of our generative model, the prior $P(\theta)$ on the parameters, is the simplest.  Our method does not place any significant constraints on the form of this probability, so one is free to choose almost any form appropriate to problem at hand, ranging from simple flat priors or factorized forms to ones that incorporate complex correlations between parameters.  The only stipulation we make is that the parameters should be continuous-valued variables (not discrete-valued), which allows for more efficient sampling procedures (see Section~\ref{sec:inference_practice}).

\begin{table*}
\setlength{\tabcolsep}{10pt}
\begin{tabular}{lll}
\hline
Model & Parameters & Data probability \\
\hline
\multirow{2}{*}{\parbox{2.3cm}{\raggedright Binomial with uniform errors}}  & True positive rate $\alpha\in[0,1]$ &
  $\mu_{ij}(1) = \alpha^{X_{ij}} (1-\alpha)^{N_{ij}-X_{ij}}$ \\
 & False positive rate $\beta\in[0,1]$ & 
  $\mu_{ij}(0) = \beta^{X_{ij}} (1-\beta)^{N_{ij}-X_{ij}}$ \\
\\
\multirow{2}{*}{\parbox{2.3cm}{\raggedright Binomial with node-dependent errors}}  & True positive rate $\alpha_i\in[0,1]$ for node~$i$ &
  $\mu_{ij}(1) = \alpha_i^{X_{ij}} (1-\alpha_i)^{N_{ij}-X_{ij}}$ \\
 & False positive rate $\beta_i\in[0,1]$ for node~$i$ & 
  $\mu_{ij}(0) = \beta_i^{X_{ij}} (1-\beta_i)^{N_{ij}-X_{ij}}$ \\
\\
\multirow{2}{*}{\parbox{2.3cm}{\raggedright Poisson with uniform errors}}  &
\multirow{2}{*}{\parbox{6cm}{\raggedright Means $\lambda_1$, $\lambda_0$ for edges and non-edges}}  &
  $\mu_{ij}(1) = \lambda_1^{X_{ij}} e^{-\lambda_1}/X_{ij}!$ \\
 & & 
  $\mu_{ij}(0) = \lambda_0^{X_{ij}} e^{-\lambda_0}/X_{ij}!$ \\
\\
\multirow{2}{*}{\parbox{2.3cm}{\raggedright Poisson with node propensity}}  &
\multirow{2}{*}{\parbox{6cm}{\raggedright Normalized node propensity $0<\eta_i<1$\\($\sum \eta_i=1$) and base rates $\lambda_1,\lambda_0$}}  &
  $\mu_{ij}(1) = (\lambda_1 \eta_i\eta_j)^{X_{ij}} e^{-\lambda_1 \eta_i\eta_j}/X_{ij}!$ \\
 & & 
  $\mu_{ij}(0) = (\lambda_0 \eta_i\eta_j)^{X_{ij}} e^{-\lambda_0 \eta_i\eta_j}/X_{ij}!$ \\
  \hline
\end{tabular}
\caption{Example data models for undirected networks with one edge type.  Here $N_{ij}$ represents the number of times the node pair~$i,j$ was measured and $X_{ij}$ represents how many of those times an edge was observed to exist.}
\label{table:data_models}
\end{table*}

\begin{table*}
\setlength{\tabcolsep}{10pt}
\begin{tabular}{lll}
\hline
Model & Parameters & Edge probability \\
\hline
Random graph & Edge probability $\rho$ & $\nu_{ij}(1) = \rho$ \\
``Soft'' configuration model &
Node pseudo-degree~$\lambda_i$ &
$\nu_{ij}(1) = 1/(1+e^{-\lambda_i\lambda_j})$ \\
\multirow{2}{*}{\parbox{4cm}{\raggedright Stochastic block model}} &
\multirow{2}{*}{\parbox{6cm}{\raggedright Node $i$ belongs to group~$g_i$ and edge probability between groups $r$ and $s$ is $\omega_{rs}$}} &
\multirow{2}{*}{\parbox{3cm}{\raggedright $\nu_{ij}(1) = \omega_{g_ig_j}$}} \\
\\
\multirow{2}{*}{\parbox{3cm}{\raggedright Random graph with multiple edge types}} &
\multirow{2}{*}{\parbox{6cm}{\raggedright Probability of type-$k$ edge $\rho_k$}} &
\multirow{2}{*}{\parbox{3cm}{\raggedright $\nu_{ij}(k) = \rho_k$}} \\
\\
\multirow{2}{*}{\parbox{3cm}{\raggedright Poisson multigraph}} &
\multirow{2}{*}{\parbox{6cm}{\raggedright Mean edge number $\omega$}} &
\multirow{2}{*}{\parbox{3cm}{\raggedright $\nu_{ij}(k) = \omega^k e^{-\omega}/k!$}} \\
\\
  \hline
\end{tabular}
\caption{Network models for the prior probability~$\nu_{ij}$ of an edge between nodes~$i$ and~$j$.}
\label{table:network_models}
\end{table*}

\subsection{Inference in theory}
\label{sec:inference_theory}
Gathering the elements defined above and substituting them into Eq.~\eqref{eq:bayes} we obtain the complete joint posterior distribution for the model:
\begin{align}
  \label{eq:joint_posterior}
  P(\bm{A},\theta|\bm{X}) &=
  \frac{P(\bm{X}|\theta,\bm{A})P(\bm{A}|\theta)P(\theta)}{P(\bm{X})},\notag\\
                          &\propto P(\theta) \prod_{(i,j)}
  \mu_{ij}(A_{ij}) \nu_{ij}(A_{ij}).
\end{align}
This distribution tells us the probability of a network structure and a set of parameter values given the observed measurements.  From it we can derive a variety of further useful quantities, such as the probability of the network structure independent of the parameters, which is given by
\begin{equation}
  P(\bm{A}|\bm{X}) = \int P(\bm{A},\theta|\bm{X})\> d\theta.
\end{equation}
Even more useful, perhaps, is the probability of having an edge of a given type between two specific nodes~$i,j$:
\begin{align}
  P(A_{ij}=k|\bm{X}) &= \int P(A_{ij}=k,\theta|\bm{X})\>d\theta\notag\\
                     &\propto \int \mu_{ij}(k,\theta)\nu_{ij}(k,\theta)P(\theta) \>d\theta,
\end{align}
where we have used Eq.~\eqref{eq:joint_posterior}.

If we instead want to learn something about a parameter $\phi\in\theta$ then we can compute its distribution as
\begin{equation}
  P(\phi|\bm{X}) = \sum_{\bm{A}} \int P(\theta',\phi,\bm{A}|\bm{X}) \>d\theta',
\end{equation}
where $\theta'$ is the parameter set with $\phi$ excluded.

Each of these quantities can be considered a special case of the posterior average of a general function~$f(\bm{A},\theta)$ of network structure and parameters, thus:
\begin{equation}
  \label{eq:general_average}
  \langle f(\bm{A},\theta) \rangle = \sum_{\bm{A}} \int f(\bm{A},\theta) P(\theta,\bm{A}|\bm{X})\ d\theta.
\end{equation}
There are a number of approaches we could take to computing expectations of this form~\cite{mclachlan2004finite}.  One possibility is to use an expectation--maximization (EM) algorithm to compute the distribution over potential networks $P(\bm{A}|\theta,\bm{X})$ as well as a point-estimate of $\theta$~\cite{newman2018estimating,newman2018network}.  Alternatively, following~\cite{peixoto2018reconstructing,peixoto2019network}, we can integrate out the parameters~$\theta$ analytically to derive the marginal distribution $P(\bm{A}|\bm{X})$ over the networks alone.  Both these approaches, however, only allow us to compute averages over network structures and not over parameters. They are moreover not in line with our goal of providing almost automatic inference for arbitrary choices of models, the EM approach because it calls for the solution of (often non-linear) equations specific to the model, and the marginal-based approach because it works only for models amenable to closed-form integration.

Instead, therefore, we employ a generalization of a method introduced in~\cite{young2019reconstruction}, which harnesses standard mixture-modeling techniques, adapting them to the network context.

\subsection{Inference in practice}
\label{sec:inference_practice}
The general idea behind our method is to compute expectations of the form~\eqref{eq:general_average} in two manageable steps by factorizing the joint posterior as
\begin{equation}
  P(\bm{A},\theta|\bm{X}) = P(\bm{A}|\theta,\bm{X})P(\theta|\bm{X}).
\end{equation}
This factorization tells us that we can draw samples from the joint posterior by first sampling sets of parameter values~$\theta$ from the marginal distribution~$P(\theta|\bm{X})$ and then sampling networks~$\bm{A}$ from $P(\bm{A}|\theta,\bm{X})$ with these parameter values.  If we sample $m$ different parameter sets and then $n$ networks for each set, we end up with $mn$ network/parameter pairs, which we number $r=1\ldots mn$.  Then we can estimate the average in Eq.~\eqref{eq:general_average} as
\begin{align}
  \langle f(\bm{A},\theta) \rangle &= \sum_{\bm{A}} \int f(\bm{A},\theta) P(\bm{A}|\theta, \bm{X}) P(\theta|\bm{X})\>d\theta \notag\\
                                   &\simeq \frac{1}{mn}\sum_{r=1}^{mn} f(\bm{A}_r,\theta_r).
  \label{eq:posterior_averages}
\end{align}
This expression is completely general and holds for any posterior, but for the class of models we consider here there are, as we now show, particularly efficient methods that can help us quickly generate the samples we need.

\subsubsection{Generating parameter samples}
The first step of the sampling algorithm draws values of the parameters~$\theta$ from the marginal distribution
\begin{equation}
P(\theta|\bm{X}) = \sum_{\bm{A}} P(\theta,\bm{A}|\bm{X}),
\end{equation}
where the sum runs over all the possible matrices~$\bm{A}$.  For models with the factorized form~\eqref{eq:joint_posterior} we have
\begin{align}
\label{eq:continuous_parameters_marginal}
P(\theta|\bm{X}) &\propto P(\theta) \sum_{\bm{A}} \prod_{(i,j)}
  \mu_{ij}(A_{ij},\theta) \nu_{ij}(A_{ij},\theta)
  \nonumber\\
 &\propto P(\theta) \prod_{(i,j)} \sum_k \mu_{ij}(k,\theta)\nu_{ij}(k,\theta).
\end{align}
Modern probabilistic programming languages make it easy to generate random samples from factorized marginals of this kind.  Our code is written in the probabilistic language Stan, which implements the technique known as Hamiltonian Monte Carlo to generate samples automatically and efficiently---see Refs.~\cite{betancourt17conceptual,carpenter2017stan} for an introduction.  Evaluating $P(\theta|\bm{X})$ involves a product over pairs $(i,j)$ of nodes, of which there are $O(n^2)$, meaning that in general generating a sample takes $O(n^2)$ time.  In many cases, however, the time complexity can be reduced to $O(n)$ by pooling terms in the product, as discussed in Sec.~\ref{subsubsec:sampling_parameters}.

\subsubsection{Generating network samples}

Given sampled values $\theta_1,\ldots,\theta_m$ of the parameters, the next step is to generate samples of the network~$\bm{A}$ from the distribution~$P(\bm{A}|\theta, \bm{X})$ for these parameter values.  This is straightforward for the factorized model assumed here, since node pairs are independent and we can sample each one separately.  Specifically, using Eqs.~\eqref{eq:joint_posterior} and \eqref{eq:continuous_parameters_marginal}, we have
\begin{align}
  P(\bm{A}|\theta, \bm{X}) &= \frac{P(\theta,\bm{A}|\bm{X})}{P(\theta|\bm{X})}
  = \frac{\prod_{(i,j)}
  \mu_{ij}(A_{ij})\nu_{ij}(A_{ij})}%
  {\prod_{(i,j)}\sum_k \mu_{ij}(k)\nu_{ij}(k)} \nonumber\\
  &= \prod_{(i,j)} Q_{ij}(A_{ij},\theta),
  \label{eq:network_cond_posterior}
\end{align}
where
\begin{equation}
  Q_{ij}(k,\theta) = \frac{ \mu_{ij}(k)\nu_{ij}(k)}{\sum_{k'} \mu_{ij}(k')\nu_{ij}(k')}
\end{equation}
is the posterior probability that nodes~$i$ and~$j$ are joined by an edge of type~$k$.  Generating networks is simply a matter of drawing a value $A_{ij}=k$ for each node pair independently from the distribution over~$k$ implied by~$Q_{ij}(k)$.  Again, naively this takes time $O(n^2)$ for all node pairs, but on a sparse network the speed can be improved by sampling only those edges with $k>0$ and assuming $k=0$ for all others.

To estimate the average $\langle f(\bm{A},\theta) \rangle$, we generate a series of parameter sets~$\theta$ using Eq.~\eqref{eq:continuous_parameters_marginal} and for each of these a series of networks using Eq.~\eqref{eq:network_cond_posterior}, then evaluate the average with Eq.~\eqref{eq:posterior_averages}.

\subsection{Assessing goodness of fit}
\label{subsec:assesing_fit}
The method described above is simple, efficient, and often gives good results.  As described in the main text, however, the method can fail if the model itself is faulty---if the model is a poor representation of the system, failing to fit the data for any parameter values.  It's important therefore to verify that the fit between model and data is good, which can be done with the standard technique of \emph{posterior-predictive assessment}.  As described in the main text, this involves generating synthetic data~$\tilde{\bm{X}}$ from the distribution implied by the fitted model:
\begin{equation}
  \label{eq:posterior_predictive}
  P(\tilde{\bm{X}}|\bm{X}) =  \int\sum_{\bm{A}} P(\tilde{\bm{X}}|\theta,\bm{A})P(\theta,\bm{A}| \bm{X})\>d\theta.
\end{equation}
This distribution weights all the possible parameters~$\theta$ and networks~$\bm{A}$ with their appropriate posterior probabilities and tells us the probability that a new data set~$\tilde{\bm{X}}$ would have if it were truly generated by the model with these inputs.  The idea of the posterior-predictive assessment is to compare these synthetic data with the original input~$\bm{X}$.  If the two look alike then the model has captured the data well; otherwise, it has not.

There are a number of ways to quantify the similarity of $\tilde{\bm{X}}$ and~$\bm{X}$.  For instance, one can compute the average
\begin{equation}
  \label{eq:posterior_predictive_mean}
  \langle \tilde{X}_{ij}\rangle = \sum_{\tilde{\bm{X}}}  P(\tilde{\bm{X}}|\bm{X}) \tilde{X}_{ij},
\end{equation}
and compare the result with~$X_{ij}$.  Visualizing the matrix of residues $\langle \tilde{\bm{X}} \rangle - \bm{X}$, the distribution of these residues, or how they depend on~$X_{ij}$ allows one to easily spot systematic issues with the model~\cite{gelman2013philosophy}.  Such calculations are not costly in practice: the distribution~\eqref{eq:posterior_predictive} is just an average of a known function of~$\bm{A},\theta$ over the posterior distribution and has the same general form as Eq.~\eqref{eq:posterior_averages}, so it can be evaluated numerically by the same methods.  In this particular case, however, we can do even better, skipping the network sampling step altogether and making an estimate directly from the parameter samples.  To do this, we write the distribution of Eq.~\eqref{eq:posterior_predictive} in the form
\begin{align}
  \label{eq:posterior_predictive_first_calculation_step}
  P(\tilde{\bm{X}}|\bm{X}) &= \int \sum_{\bm{A}} P(\tilde{\bm{X}}|\theta,\bm{A})P(\bm{A}|\theta,\bm{X}) P(\theta|\bm{X})\>d\theta \nonumber\\
  &= \int P(\theta|\bm{X}) \sum_{\bm{A}} \prod_{(i,j)} \tilde{\mu}_{ij}(A_{ij},\theta)\,Q_{ij}(A_{ij},\theta) \>d\theta \nonumber\\
  &= \int P(\theta|\bm{X}) \prod_{(i,j)} \sum_k \tilde{\mu}_{ij}(k,\theta)\,Q_{ij}(k,\theta) \>d\theta,
\end{align}
where have used Eqs.~\eqref{eq:data_model} and~\eqref{eq:network_cond_posterior} in the second line, and $\tilde{\mu}_{ij}(k,\theta)$ is the probability of generating a synthetic measurement~$\tilde{X}_{ij}$ given that $(i,j)$ is an edge of type~$k$.  This expression is now independent of~$\bm{A}$ and only requires an average over~$\theta$ to evaluate.

Using this expression for $P(\tilde{\bm{X}}|\bm{X})$, we can write the average $\langle \tilde{X}_{ij}\rangle$ in Eq.~\eqref{eq:posterior_predictive_mean} as
\begin{equation}
  \langle \tilde{X}_{ij}\rangle = \int P(\theta|\bm{X}) \sum_k
\bigl\langle \tilde{\mu}_{ij}(k,\theta) \bigr\rangle\,Q_{ij}(k,\theta)\>d\theta,
\end{equation}
which we evaluate numerically as
\begin{equation}
  \langle \tilde{X}_{ij}\rangle  \simeq \frac{1}{m} \sum_{r=1}^m\sum_k \bigl\langle \tilde{\mu}_{ij}(k,\theta_r) \bigr\rangle\,Q_{ij}(k,\theta_r).
\label{eq:avtildeX}
\end{equation}
Note that $\bigl\langle \tilde{\mu}_{ij}(k,\theta_r) \bigr\rangle$ usually has a simple closed form, since it is just the mean of $\tilde{X}_{ij}$ within the data model with parameters~$\theta_r$.

A visual inspection of the residues between $\tilde{\bm{X}}$ and~$\bm{X}$ is often enough to reveal issues with goodness of fit, but one can carry out a more formal model assessment using any of a variety of \emph{discrepancy} measures that quantify the distance between the synthetic data~$\tilde{\bm{X}}$ and the original~$\bm{X}$~\cite{gelman1996posterior}.  The average value of such a discrepancy will always be greater than zero, since one does not expect the synthetic and original data to agree perfectly even with a perfect model.  To obtain a baseline against which discrepancy values can be compared, we therefore compute the discrepancy between synthetic measurements~$\tilde{\bm{X}}$ and \emph{their} associated predictions, calculating a model-versus-model discrepancy distribution.

In the calculations presented here we make use of the log-likelihood ratio discrepancy:
\begin{equation}
  \label{eq:discrepancy}
  D(\bm{X},\theta_r) =  \sum_{(i,j)} X_{ij} \log \frac{X_{ij}}{\tilde{X}_{ij}(\theta_r)},
\end{equation}
where $\tilde{X}_{ij}(\theta_r)$ is evaluated using Eq.~\eqref{eq:avtildeX} with the sampled parameter values~$\theta_r$.  This discrepancy is reminiscent of a Kullback-Leibler divergence, with the primary difference being that it compares unnormalized quantities rather than normalized probability distributions.  That said, the norm of the two sets of measurements should be similar, since the whole purpose of the calculation is to reproduce the original observations.  Hence, one can usually interpret the discrepancy in more or less the same way as a divergence: the smaller the divergence the better the fit (although values slightly less than zero can occur, which is not true of a true divergence).

We compute the distribution of the discrepancy and the reference distribution~$\tilde{\bm{X}}$ simultaneously using the method introduced in Ref.~\cite{gelman1996posterior}.  We go through each network/parameter sample $\bm{A}_r,\theta_r$ and generate a single realization~$\tilde{\bm{X}}$ of the synthetic data from the data model, then compute the two discrepancies $D(\tilde{\bm{X}},\theta_r)$ and $D(\bm{X},\theta_r)$ using the analog of Eq.~\eqref{eq:avtildeX}.  From the resulting sets of discrepancy values one can then compute the $p$-value $p = P[D(\bm{X},\theta_r) > D(\tilde{\bm{X}},\theta_r)]$, which is the fraction of artificial data sets with discrepancy at least as large as the observed value.  The model is rejected if the $p$-value is too small.  This calculation does not cost much computation time since we are merely reusing the samples already generated for estimation purposes.

\subsection{Implementation}
In this section we discuss details of implementation of the algorithm, including a number of techniques for improving speed and numerical accuracy which can be useful with large data sets.

\subsubsection{Sampling networks}
One of the more computationally costly steps in the algorithm is the generation of sample networks from the conditional posterior distribution~$P(\bm{A}|\theta,\bm{X})$.  Naively generating the network by flipping a biased coin for every node pair~$i,j$ takes time~$O(n^2)$ on a network of~$n$ nodes.

For some models on sparse networks this time can be reduced by explicitly sampling only the edges that exist.  That is, all edges are assumed not to exist, except for a sparse sample that are generated in accordance with the fitted model.  For instance, with the simple ``uniform error'' model of Table~\ref{table:data_models}, the posterior probabilities~$Q_{ij}$ of edges are a unique function $Q(X)$ of the number of observations~$X_{ij}$ of the edge in question.  With this in mind we define $\Sigma = \sum_{(i,j)} Q_{ij} = \sum_X n(X) Q(X)$ where $n(X) = \sum_{(i,j)} \delta(X, X_{ij})$ is the number of node pairs with $X$ observations and $\delta(x,y)$ is the Kronecker delta.

The value of $\Sigma$ can be calculated rapidly once $n(X)$ is known, then we can generate a sampled network by first drawing an integer~$M \sim \mbox{Poisson}(\Sigma)$ to represent the number of edges in the network, and then generating $M$ random edges with probabilities~$Q_{ij}$ with standard ``roulette wheel'' proportional sampling using binary search.  The complete process takes time~$O(M \log n)$, which on a sparse network will be much faster than the $O(n^2)$ of the naive algorithm.

In other cases we may be able to skip the process of network sampling altogether, although at the price of still having to perform $O(n^2)$ operations.  Specifically, when we want to calculate the average of a function~$f$ that factorizes over node pairs thus
\begin{equation}
  \label{eq:factorizable_functions}
  f(\bm{A},\theta) = \prod_{(i,j)} g_{ij}(A_{ij},\theta),
\end{equation}
we can write the average as
\begin{align}
  &\langle f(\bm{A},\theta)\rangle = \sum_{\bm{A}} \int f(\bm{A},\theta) P(\bm{A}|\theta, \bm{X}) P(\theta|\bm{X})\>d\theta \notag\\
  &\qquad{} =  \int   P(\theta|\bm{X}) \prod_{(i,j)}\sum_k  \left[g_{ij}(k,\theta) Q_{ij}(k,\theta) \right] \>d\theta.
\end{align}
Now we sample $m$ sets of parameter values~$\theta_r$ as usual, but generate no networks~$\bm{A}$, and the average we want is given by
\begin{equation}
  \langle f(\bm{A},\theta) \rangle \simeq \frac{1}{m}\sum_{r=1}^m\,
  \prod_{(i,j)} \sum_k g_{ij}(k,\theta_r) Q_{ij}(k,\theta_r) .
  \label{eq:average_over_factorizable_functions}
\end{equation}

\subsubsection{Sampling parameters}
\label{subsubsec:sampling_parameters}
Generating sample values of the parameters also takes time~$O(n^2)$ in general, because the right-hand side of Eq.~\eqref{eq:continuous_parameters_marginal} involves a product over pairs of nodes.  For some models, however, we may be able to evaluate this product more rapidly by methods similar to those described for sampling networks above.  Taking again the example of the ``uniform error'' model from Table~\ref{table:data_models}, the probability $\mu_{ij}(k)$ is a function~$\mu(X,k,\theta)$ only of the number of observations~$X_{ij}$ of the corresponding edge (and~$k$ and~$\theta$) and $\nu_{ij}(k)$ is a function of~$k$ and~$\theta$ only.  This means we can group terms in the product and write
\begin{equation}
\prod_{(i,j)} \sum_k \mu_{ij}(k,\theta)\nu_{ij}(k,\theta)
  = \prod_X \biggl[ \sum_k \mu(X,k,\theta) \nu(k,\theta) \biggr]^{n(X)},
\end{equation}
which saves considerable time.

\end{document}